\documentclass[12pt]{article}
\usepackage{putex}
\usepackage[pdftex]{graphicx}

\begin{document}

\preprint{COLO-HEP-559 \\ PUPT-2360}

\institution{CU}{${}^1$Department of Physics, 390 UCB, University of Colorado, Boulder, CO 80309, USA}
\institution{PU}{${}^2$Joseph Henry Laboratories, Princeton University, Princeton, NJ 08544, USA}

\title{A holographic critical point}

\authors{Oliver DeWolfe,${}^\CU$ Steven S.~Gubser,${}^\PU$ and Christopher Rosen${}^\CU$}

\abstract{We numerically construct a family of five-dimensional black holes exhibiting a line of first-order phase transitions terminating at a critical point at finite chemical potential and temperature.  These black holes are constructed so that the equation of state and baryon susceptibilities approximately match QCD lattice data at vanishing chemical potential.  The critical endpoint in the particular model we consider has temperature 143 MeV and chemical potential 783 MeV.  Critical exponents are calculated, with results that are consistent with mean-field scaling relations.}

\date{December 2010}

\maketitle

\section{Introduction}
\label{INTRODUCTION}

At zero chemical potential $\mu$ for baryon number, quantum chromodynamics (QCD) appears to have a smooth but rapid crossover at a temperature $T_c$ whose value is within about $10\%$ of $175\,{\rm MeV}$.  It is believed that this crossover sharpens into a line of first order phase transitions at finite $\mu$.  The position of the critical point that terminates this line is of considerable experimental interest, but it is hard to determine theoretically due to being in a region of strong coupling, and also because lattice techniques are not well adapted to finite real $\mu$.  A theory review can be found in \cite{Rischke:2003mt}.  Recent lattice results can be found in for example  \cite{Karsch:2007dp,Bazavov:2009zn,Borsanyi:2010cj}.  The aims of the recently initiated beam-energy scan at RHIC are laid out in \cite{Aggarwal:2010cw} and the fixed-target CBM project at FAIR is discussed in \cite{Staszel:2010zz}.

It was shown in \cite{Gubser:2008ny,Gubser:2008yx} that simple gravitational theories in five dimensions are capable of producing black holes which approximately reproduce the equation of state of QCD, including the crossover, at vanishing chemical potential; see also \cite{Gursoy:2008bu,Gursoy:2008za,Gursoy:2009jd,Noronha:2009ud,Charmousis:2010zz} and the review \cite{Gursoy:2010fj}.  The gravitational theories include just two fields: the spacetime metric, and a real scalar field whose profile breaks conformal invariance and can be understood roughly as the running coupling of QCD.  A natural generalization of such models is to include a chemical potential for baryon number as well.  Adding a single additional field, a $U(1)$ gauge field dual to the baryon number current, one may generate a chemical potential by turning on an appropriate electric field in the black hole geometry.

In this paper, we study the coupled metric-scalar-gauge field system and numerically obtain solutions for charged black holes filling out the $T$-$\mu$ phase diagram of the field theory dual.  The minimal Lagrangian for these theories contains some freedom, encoded in the choice of scalar potential and gauge kinetic function.  We elect to fix both these functions by matching to lattice results for QCD at zero chemical potential.  The scalar potential is fixed by demanding a QCD-like equation of state that captures the chiral symmetry breaking crossover, as in \cite{Gubser:2008ny,Gubser:2008yx}.  We show how the gauge kinetic function can be determined by similarly matching quark susceptibilities, in principle removing all freedom from the construction.

We then investigate how black holes behave at finite chemical potential.  We find that just as is expected for QCD, at finite $\mu$ the crossover turns into a line of true first-order phase transitions ending in a critical point.  We locate the first-order line by looking for characteristic thermodynamically unstable solutions, and identify the critical point as the end of this line; the location of the critical point is at physically reasonable values of $T$ and $\mu$.

We then turn to a study of the critical exponents of this point.  We find a set of exponents that are nontrivially self-consistent due to satisfying two scaling relations.  Thus our black holes built from just three fields reproduce a realistic phase diagram near the critical endpoint for a QCD-like theory.  In QCD, the critical point is expected to lie in the same universality class as the 3D Ising model and the fluid liquid/gas transition.   The critical exponents we obtain are consistent with mean-field scaling.  This is reasonable since our black hole constructions are classical, corresponding to an implicit large $N$ limit on the field theory side that suppresses quantum corrections.  Further realism lies, presumably, in the inclusion of $1/N$ corrections.

The organization of the rest of this paper is as follows.  In section~\ref{SUMMARY} we describe our gravity theory and summarize our results for the location of the critical point in the $T$-$\mu$ plane and the values of its critical exponents. In section~\ref{GENERAL} we provide a self-contained summary of the aspects of thermodynamics which we will require in the rest of the paper, as well as brief remarks on the phase structure of QCD.  Experts will have no reason to read this section, which contains no new results.  In section~\ref{EOMS} we analyze the equations of motion following from our gravity theory, 
explain how to extract thermodynamic quantities from the black hole solutions, and summarize our numerical strategy.
 In section~\ref{SUSCEPTIBILITY} we will explain how  the gauge kinetic function can be chosen to match lattice data for the baryon susceptibility at $\mu=0$.  
In section~\ref{SEARCH} we describe locating the critical point on the phase diagram, and in 
section~\ref{ANALYSIS} we analyze its properties, calculating the critical exponents and finding them consistent with mean-field scaling.
In section~\ref{CONCLUSIONS} we compare our result for the location of the critical point to others in the literature, and conclude with some discussion.

\section{Gravity Theory and Summary of Results}
\label{SUMMARY}

Our model falls in a class of  five-dimensional gravitational theories including a real scalar $\phi$ and an abelian gauge field $A_\mu$ along with the spacetime metric, defined by the Lagrangian
\eqn{LwithF}{
  {\cal L} = {1 \over 2\kappa^2} \left[ 
    R - {f(\phi) \over 4} F_{\mu\nu}^2 - {1 \over 2} (\partial\phi)^2 - V(\phi) \right] \,,
 }
where we use mostly plus conventions.  
 With energy dimensions assigned so that $[\kappa]=-3/2$, $[g_{\mu\nu}]=[A_\mu]=[\phi]=0$, one finds that \eno{LwithF} 
is almost\footnote{The exception is that one could add a Chern-Simons term $A \wedge F \wedge F$, but this term has no effect on 
the classical equations of motion, which are what we will study; in addition it vanishes for
the solutions we are going to consider, because they have only electric charge, not magnetic.  Thus we neglect it. } the most general action using this field content one can have with at most two derivatives.      Arbitrary functions of $\phi$ multiplying the Einstein-Hilbert and scalar kinetic terms can be removed by conformal transformations of the metric and reparametrizations of $\phi$, respectively.

The black hole geometries consist of metrics taking the form
 \eqn{BHgeom}{
  ds^2 = e^{2A(r)} \left[ -h(r) dt^2 + d\vec{x}^2 \right] + {e^{2B(r)} \over h(r)} dr^2 \,,
 }
 along with an ansatz for the scalar field and electrostatic potential depending only on the radial coordinate $r$:
 \eqn{ScalarAndVector}{
 \phi = \phi(r) \,, \quad \quad A_\mu dx^\mu = \Phi(r) dt \,.
 }
The coordinates $(t,\vec{x})$ cover Minkowski space, ${\bf R}^{3,1}$, while the radial coordinate $r$ represents the holographic direction.

As explained in \cite{Gubser:2008ny} in the case of vanishing gauge field, a choice of the scalar potential $V(\phi)$ can be translated into a dependence of the entropy on temperature $T$.  (\cite{Gubser:2008ny} chooses to work equivalently with the speed of sound $c_s^2 = d \log T / d \log s$.)  In fact, if a desired dependence $s(T)$ is specified, then---within certain limits---one can find the $V(\phi)$ that leads to it.  A reasonable fit, not too far from $T_c$, to lattice results for $s(T)$, is achieved with the simple choice\footnote{The constant $\gamma$ in \eno{VChoice} is unrelated to the critical exponent which will appear later in the paper.} \cite{Gubser:2008yx}
 \eqn{VChoice}{
  V(\phi) = {-12 \cosh \gamma\phi + b\phi^2 \over L^2} \qquad
    \hbox{with $\gamma = 0.606$ and $b = 2.057$} \,,
 }
 and $L$ a constant related to the number of degrees of freedom.
 
The black holes describing matter at finite chemical potential include a nonzero gauge field as well.  This introduces the problem of specifying the gauge kinetic function $f(\phi)$.  We can always use the freedom to rescale $A_\mu$ to set $f(0) = 1$.
The matching of the speed of sound described in the last paragraph is completely insensitive to the choice of $f(\phi)$.
However, $f(\phi)$ can be fixed if one knows the baryon number susceptibility at $\mu=0$.  This susceptibility is in fact fairly well known from the lattice \cite{Karsch:2007dp}.  
In this paper, we will not be systematic in finding $V(\phi)$ and $f(\phi)$ through a fit.  Instead, we will focus on the above choice of $V(\phi)$ and a similarly simple form for $f(\phi)$, namely
 \eqn{fChoice}{
  f(\phi) = {\sech\left[ {6 \over 5} (\phi-2) \right] \over \sech {12 \over 5}} \,,
 }
which as we will discuss in section~\ref{SUSCEPTIBILITY} leads to susceptibilities in good agreement with lattice results.

It is probably impossible to find a string theory construction that leads precisely to the potential \eno{VChoice} and gauge kinetic function \eno{fChoice}.  Thus we cannot claim that the theory \eno{LwithF} is dual to a specific known field theory.  However, string theory constructions do typically lead to potentials which include sums of exponentials of canonically normalized scalars.  Thus these functions are at least in the ballpark of expressions that can be derived from string theory, and it is reasonable to place the dual to our model in the broad class of strongly coupled, large-$N$ gauge theories.

Having fit $V(\phi)$ and $f(\phi)$ to lattice quantities at $\mu=0$, we are able to use black hole constructions to extrapolate outward into the $T$-$\mu$ plane, where we indeed find a critical endpoint. Through methods explained in sections~\ref{SUSCEPTIBILITY} and~\ref{SEARCH}, we estimate the location of this critical point to be
 \eqn{CriticalPosition}{
  T_c = 143 \ {\rm MeV} \qquad \mu_c = 783 \ {\rm MeV} \,.
 }
 We will compare this result to other estimates in the literature in the conclusions.
 
Because we have not made a systematic study of the forms of $V(\phi)$ and $f(\phi)$ that approximately match lattice data at $\mu=0$, we are not in a position to provide theoretical error bars for the result \eno{CriticalPosition}.  It is best to view this result as a 
proof of principle that you can get a critical endpoint in the $T$-$\mu$ plane using AdS/CFT methods, and that the values \eno{CriticalPosition} are within the theoretical error bars.
It is also noteworthy that we ignore fluctuations in our analysis: the black holes we construct are fixed, classical geometries.  This means that we are not capturing all the physics that is expected to go into the critical endpoint.

Analyzing the thermodynamics near the critical point and performing linear regression fits to the data, we obtain results for four critical exponents,
\eqn{CritExpSummary}{
\alpha = 0 \,, \quad \quad
\beta \approx 0.482 \,, \quad \quad
\gamma \approx 0.942 \,, \quad \quad 
\delta \approx  3.035 \,. 
}
These results are, as we shall discuss, non-trivially consistent with scaling relations, and consistent also with the mean field exponents $\alpha =0$, $\beta = 1/2$, $\gamma = 1$, $\delta = 3$.

\section{Thermodynamics with a finite chemical potential}
\label{GENERAL}

Before discussing the solution of the equations of motion in the gravity system \eno{LwithF} and the exploration of the phase diagram, in this section we review a few essential aspects of thermodynamics and critical phenomena.

\subsection{Thermodynamics of a fluid}

A fluid is characterized by the extensive quantities entropy $S$, volume $V$ and particle number (or net charge) $N$, and their conjugate intensive variables temperature $T$, pressure $p$ and chemical potential $\mu$.  The internal energy $U = U(S, V, N)$ depends on the extensive variables, which can be thought of as characterizing the system itself, with small changes described by the first law,
\eqn{FirstLaw}{
dU = TdS - pdV + \mu dN \,.
}
The intensive variables, sometimes called ``fields," can be thought of as properties imposed on the system by contact with a reservoir.

For systems like the quark-gluon plasma, we are not interested in a fixed volume, but instead in volume densities for the extensive quantities.
Define the energy density $\epsilon$, entropy density $s$ and number density $\rho$, 
\eqn{DefineDensities}{
\epsilon \equiv U/V \,, \quad \quad s \equiv S/V \,, \quad \quad \rho \equiv N/V \,.
}
Then, using the thermodynamic relation,
\eqn{ThermRel}{
U = TS - pV + \mu N\,,
}
one can show that the first law of thermodynamics \eno{FirstLaw} rewritten in terms of densities becomes
\eqn{DensityFirstLaw}{
d\epsilon = T ds + \mu d\rho \,,
}
naturally reducing the system to a two-variable problem.  It is useful to define the corresponding free energy density $f(T, \mu)$ depending on the field variables $T$ and $\mu$ using the usual Legendre transformation,\footnote{The free energy density should not be confused with the function $f(\phi)$ in the gravity Lagrangian.}
\eqn{FreeEnergy}{
f(T, \mu) \equiv \epsilon - sT - \mu \rho \,,
}
which obeys
\eqn{}{
df = - s dT  - \rho d \mu \,.
}
Furthermore, it is easy to see the thermodynamic relation \eno{ThermRel} implies that the pressure reappears in the analysis as just minus the free energy,
\eqn{PressureFreeEnergy}{
p = -f \,.
}
The phase diagram is the plot of ``field" variables $T$ and $\mu$.
At each point on the phase diagram, a physical phase corresponds to values of the extensive variables (or  densities of extensive variables in our case, $s$ and $\rho$) which extremize the free energy.  In general, it is possible for more than one extremum to exist at a given point on the diagram, corresponding to the existence of multiple phases.  The preferred phase is the one minimizing the free energy $f$; this is the condition of global stability and ultimately stems from the second law of thermodynamics.

In addition to global stability, one must consider local stability, which is characterized by stability under small fluctuations.  This is equivalent to the statement of positive-definiteness of the matrix of susceptibilities:
\eqn{SusceptMatrix}{
{\cal S} \equiv \begin{pmatrix} -{\partial^2 f \over \partial T^2} & -{\partial^2 f \over \partial \mu \partial T} \\
   -{\partial^2 f \over \partial T \partial \mu} &-{\partial^2 f \over \partial \mu^2} \end{pmatrix} 
   = 
    \begin{pmatrix} {\partial s \over \partial T} & {\partial s \over\partial \mu} \\
   {\partial \rho \over \partial T} &{\partial\rho \over \partial \mu} \end{pmatrix} \,,
}
where all derivatives of $T$ or $\mu$ are taken with the other fixed.  We note that the upper-left diagonal element is related to the specific heat at constant chemical potential $C_\mu$:
\eqn{CMuEqn}{
C_\mu \equiv T \left( \partial s \over \partial T \right)_\mu = - T \left( \partial^2 f \over \partial T^2 \right)_\mu \,,
}
while the lower-right diagonal quantity is related to the isothermal compressibility,
 \eqn{kappaTDef}{
  \kappa_T \equiv {1 \over \rho^2} \left( {\partial\rho \over \partial\mu} \right)_T \,.
 }
In QCD, one usually considers in place of $\kappa_T$ the quark susceptibility,
\eqn{ChiEqn}{
\chi_2 \equiv \left( \partial \rho \over \partial \mu \right)_T = - \left( \partial^2 f \over \partial \mu^2 \right)_T = \rho^2 \kappa_T \,.
}
The statement that the matrix of susceptibilities is positive definite is equivalent to requiring
\eqn{LocalStab}{
C_\rho > 0 \,, \quad \quad \chi_2 > 0 \,, 
}
where $C_\rho$ is the specific heat at constant volume,
\eqn{CRhoEqn}{
C_\rho \equiv T \left( \partial s \over \partial T \right)_\rho = - T \left[ { \partial^2 f \over \partial T^2 } -  {\left( \partial^2 f / \partial T \partial \mu\right)^2   \over ( \partial^2 f /\partial \mu^2)} \right] \,.
}
In the last expression all $T$ and $\mu$ derivatives are taken with the other fixed.  Note that the Jacobian of the susceptibility matrix is simply
\eqn{JacSuscept}{
\det {\cal S} = {1 \over T} \chi_2 C_\rho\,.
}
Generally speaking, local stability (i.e.~positive definite ${\cal S}$) is a requirement in order for a configuration to be considered a well-defined phase of the system.

When two phases have equal free energies at a given $(T, \mu)$, a first-order phase transition occurs at that point; typically the locus of first-order transitions has codimension one and so describes a line.  On one side of the line one phase is favored, while on the other the other is favored; at the first-order line, a discontinuity exists in the densities, $\Delta s$ and $\Delta \rho$, as the system jumps from one phase to another.  The discontinuity in the entropy gives rise to the latent heat $L = T \Delta S$.

A first-order line may terminate on a second-order point, also called a critical point.  Here the two distinct phases merge into one; consequently the discontinuities in the densities approach zero as one moves along the first-order line to the critical point.  However, although these first derivatives of the free energy become well-behaved, the second derivatives, that is the susceptibilities, may diverge at the critical point.  The nature of these divergences characterize some of the critical exponents associated to the critical point, as we will describe momentarily.  Critical points in different systems with different variables may share the same critical exponents, a phenomenon called ``universality;'' the systems are said to lie in the same universality class.  Beyond the critical endpoint on the phase diagram there is no discontinuous behavior, but the densities may change rapidly along the extension of what would have been the first-order line.  This behavior is called a crossover.

\subsection{Phase diagram for QCD}

A great deal of work has gone into predicting the phase structure of QCD, and it is believed to be quite rich; for reviews, see \cite{Kogut:2004su,Stephanov:2007fk, Alford:2007xm}.  At small values of the baryon chemical potential, the QCD phase diagram is dominated by the chiral symmetry breaking transition, and this will be our focus.

When all quarks are assumed to be massless, chiral symmetry is an exact symmetry of the QCD Lagrangian, and the broken symmetry phase at low $T$ and $\mu$ and the restored symmetry phase at high $T$ and/or $\mu$ are distinct and must be separated by a line of true phase transitions.  Near the $\mu$-axis, the transition is expected to be first-order.  Near the $T$-axis, the order of the transition depends on the number of massless quarks.  For two massless quarks, the transition on the $T$-axis is second order and in the universality class of the $O(4)$ model; this transition is expected to be the end of a line of second-order transitions extending into the $T$-$\mu$ plane and meeting the first-order line rising from the $\mu$-axis at a tricritical point.  For three massless quarks, on the other hand, the transition on the $T$-axis is expected to be first-order.

\begin{figure}
\begin{center}
\includegraphics[scale=0.35]{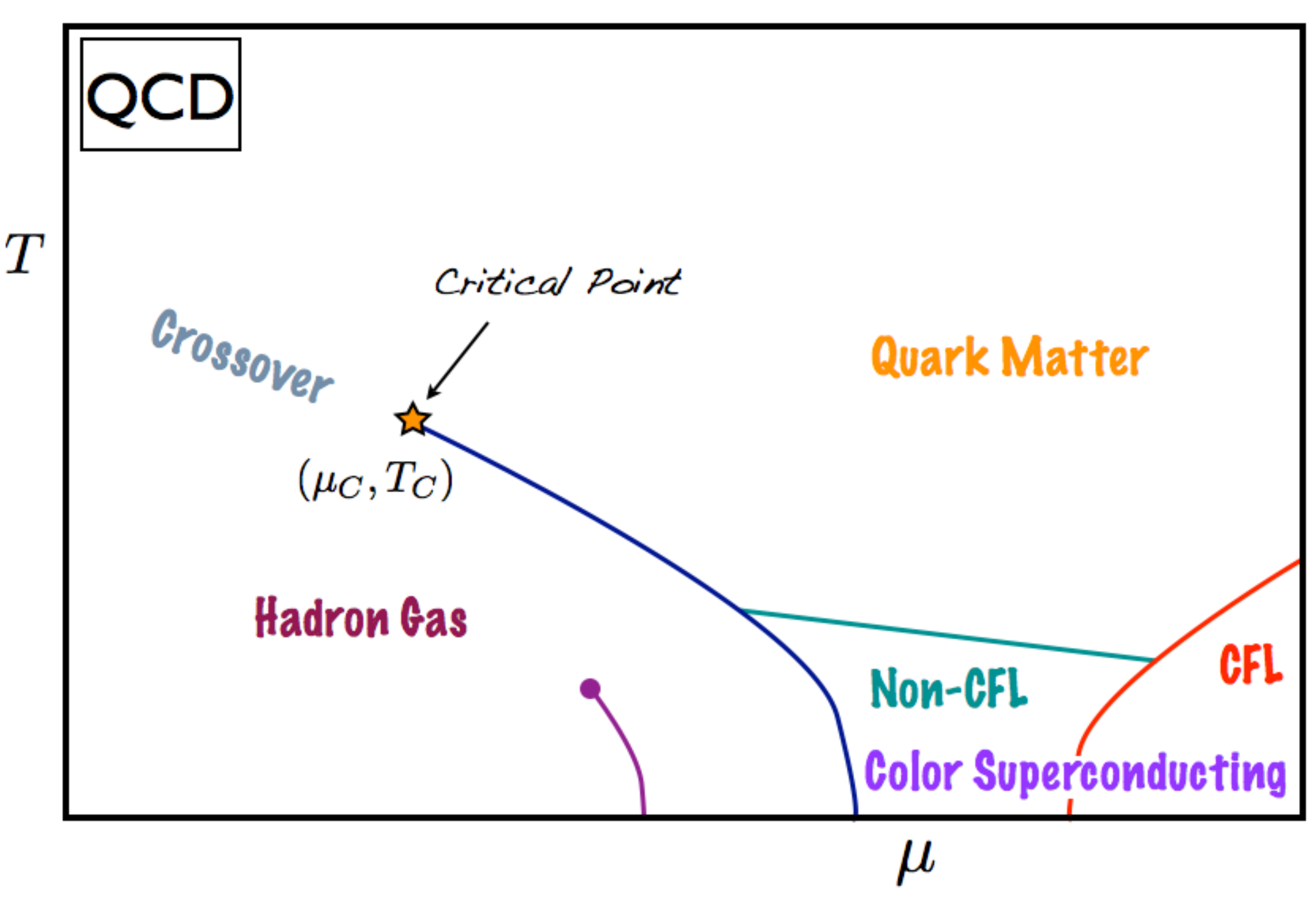}
\caption{The expected phase diagram of QCD.  The line ending in a star is the first-order chiral transition and its critical endpoint, which we focus on. Below is the nuclear matter transition.  At lower right are color superconducting phases, color-flavor locked and otherwise.
\label{fig:PhaseDiag}}
\end{center}
\end{figure}

In the real world, quarks are massive and chiral symmetry is not an exact symmetry of QCD.  On the $T$-axis, the transition is known from lattice studies not to be a sharp transition but instead a crossover.  It is widely expected that at sufficiently large chemical potential $\mu$ the first-order line returns; it then terminates at a critical endpoint at some $(T_c, \mu_c)$.  This is displayed in figure~\ref{fig:PhaseDiag}.

The critical endpoint is an object of substantial interest and speculation.  It is difficult to explore it theoretically, as the theory is strongly coupled and lattice calculations are difficult at finite $\mu$.  A number of models have been constructed to analyze its properties.  It is expected to lie in the universality class of the 3D Ising model, like the standard liquid/gas transition of fluids.
It is anticipated that depending on its location on the phase diagram, future heavy ion experiments such as those at RHIC, LHC or FAIR may produce a quark-gluon plasma lying close to the critical point at freeze-out, which could lead to information about its properties (see for example \cite{Aggarwal:2010cw,Staszel:2010zz, Stephanov:1998dy}.)  

Other phases of QCD are anticipated to exist, in particular regions at large $\mu$ characterized by color superconductivity.  We will have little to say about these phases in this paper other than a brief speculation in section~\ref{LOCATING}.

\subsection{Critical behavior}

Near the critical point various first and second derivatives of the free energy go to zero or diverge as power laws, and  it is the  ``critical exponents'' associated with these power laws that are universal --- meaning they may take the same values from one physical system to another, even among systems with quite different microscopic properties.  Describing them is at the heart of the study of critical phenomena.
 
\begin{figure}
  \centerline{\includegraphics[width=4in]{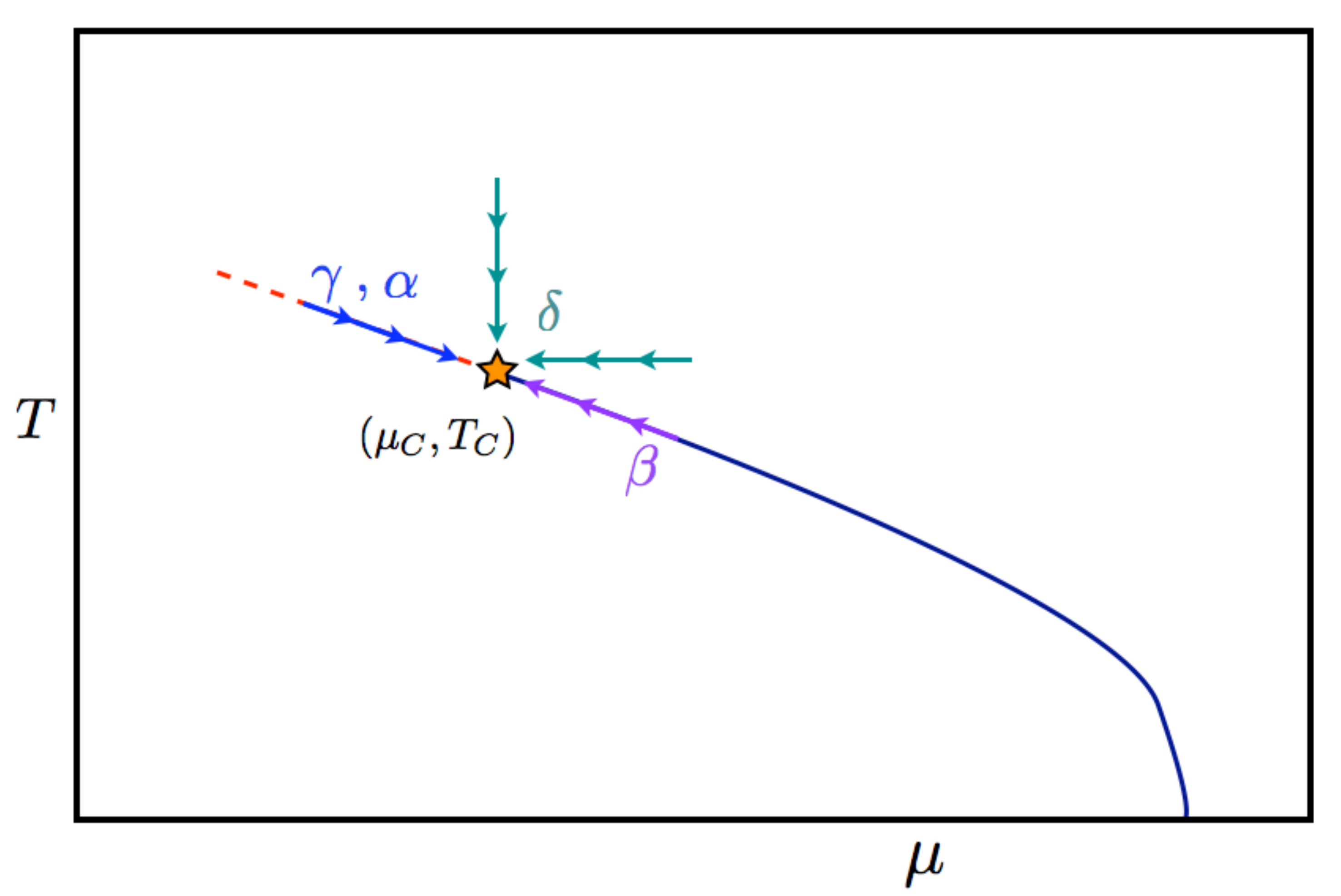}}
    \caption{A cartoon of the first-order line terminating at the critical point (star) with the directions of approach of the various critical exponents indicated.}
    \label{critExp}
 \end{figure}

We will calculate four standard thermodynamic critical exponents $\alpha$, $\beta$, $\gamma$ and $\delta$.\footnote{Two other commonly-used critical exponents, $\nu$ and $\eta$, require knowledge of the spatial distribution of correlation functions and will not be calculated here.}  In calculating the exponents, it is vital to specify whether one is approaching the critical point along the axis defined by the first order line, or by another direction.  
The exponent $\alpha$ is defined by the power law behavior of the specific heat at constant $\rho$ as the critical point is approached along the axis defined by the first order line:
\eqn{AlphaDef}{
C_\rho \sim |T - T_c|^{-\alpha} \,, \quad \quad \quad {\rm along \; first \; order \; axis} \,.
}
The exponent $\beta$ comes from the discontinuity  of $\rho$ across the first-order line.  $\Delta \rho$ is finite at a generic point on the first-order line, and goes to zero as one approaches the critical point along the line:
\eqn{BetaDef}{
\Delta \rho \sim (T_c - T)^\beta \,, \quad \quad \quad {\rm along \; first \; order \; line} \,.
}
The exponent $\gamma$ is analogous to $\alpha$, but instead of $C_\rho$, it is $\chi_2$ that is tracked along the first-order axis:
\eqn{GammaDef}{
\chi_2 \sim |T - T_c|^{-\gamma} \,, \quad \quad \quad {\rm along \; first \; order \; axis} \,.
}
Finally, $\delta$  is defined at the critical isotherm $T = T_c$  by the relation between $\rho - \rho_c$ and $\mu- \mu_c$:
\eqn{DeltaDef}{
\rho - \rho_c \sim |\mu - \mu_c|^{1/\delta} \,, \quad \quad \quad {\rm for}\ T = T_c \,.
}
The same power law will manifest for any approach {\em not} parallel to the first-order line.
The paths through the phase diagram associated to the four exponents are summarized in figure~\ref{critExp}. 

The four thermodynamic exponents are not all independent; in general they obey so-called scaling relations, which follow from the scaling behavior of the free energy at the critical point, determining two exponents in terms of the other two.  One has
\eqn{Scaling}{
\alpha + 2\beta + \gamma  &= 2  \,, \cr
\alpha+ \beta(1+\delta) &= 2\,.
}
Different critical exponents are characteristic of distinct universality classes.
Calculations in Landau-Ginzburg, or mean-field, theory capture the tree-level values of the critical exponents; in general this neglects quantum corrections, which can be captured by the more sophisticated techniques of the renormalization group.  The critical point of QCD is expected to lie in the universality class of the 3D Ising model, as does the standard liquid/gas transition.  The results from mean field (van der Waals) theory, the full quantum 3D Ising model, and experiments in non-QCD fluids are summarized in the table \cite{Goldenfeld}:

\begin{center}
\begin{tabular}{|c|ccc|}
\hline
& Mean field  & 3D Ising& Experiment \cr  \hline
$\alpha$ & 0 & 0.110(5) & 0.110 - 0.116 \cr
$\beta$ & 1/2 & 0.325 $\pm$ 0.0015 & 0.316 - 0.327\cr
$\gamma$ & 1 & 1.2405 $\pm$ 0.0015 & 1.23 - 1.25 \cr
$\delta$ & 3  & 4.82(4) &4.6 - 4.9\cr
\hline
\end{tabular}
\end{center}
These are the results we will compare our holographic system to.

\section{Black hole solutions}
\label{EOMS}

We now turn to an analysis of the equations of motion for the gravity system.
From the action \eno{LwithF} one can derive four second order equations of motion and a zero-energy constraint.  The second order equations (simplified slightly using the zero-energy constraint) are
 \eqn{SecondOrder}{
  A'' - A' B' + {1 \over 6} \phi'^2 &= 0  \cr
  h'' + (4A'-B') h' - e^{-2A} f(\phi) \Phi'^2 &= 0  \cr
  \Phi'' + (2A' - B') \Phi' + {d\log f \over d\phi} \phi' \Phi' &= 0  \cr
  \phi'' + \left( 4A' - B' + {h' \over h} \right) \phi' - {e^{2B} \over h} 
    {\partial V_{\rm eff} \over \partial\phi} &= 0 \,,
 }
where
 \eqn{VeffDef}{
  V_{\rm eff}(\phi,r) \equiv V(\phi) - {1 \over 2} e^{-2A-2B} f(\phi) \Phi'^2 \,.
 }
The zero-energy constraint is
 \eqn{ZeroEnergy}{
  h (24 A'^2 - \phi'^2) + 6 A' h' + 2 e^{2B} V(\phi) + e^{-2A} f(\phi) \Phi'^2= 0 \,.
 }
The equation of motion for $\Phi$ can be integrated once to show 
the conservation of the Gauss charge $Q_G$ for the $U(1)$ gauge field:
 \eqn{GaussConstraint}{
  {dQ_G \over dr} = 0 \qquad\hbox{where}\qquad Q_G = f(\phi) e^{2A-B} \Phi' \,.
 }
One other conserved quantity can be guessed from scaling symmetries, as in \cite{Gubser:2008ny}:
 \eqn{NoetherConstraint}{
  {dQ_N \over dr} = 0 \qquad\hbox{where}\qquad Q_N = e^{2A-B} [e^{2A} h' - f(\phi) \Phi \Phi'] \,.
 }

\subsection{Near-horizon asymptotics}
\label{NEAR}

Let's assume that $h$ has a simple zero at $r_H$ and that it has no additional zeroes between $r_H$ and the boundary.  Then $r_H$ is the location of a regular black hole horizon.  A series solution to the equations \eno{SecondOrder} and \eno{ZeroEnergy} can be developed simply by expanding
 \eqn{Xexpand}{
  X(r) = X_0 + X_1 (r-r_H) + X_2 (r-r_H)^2 + \ldots \,,
 }
where $X$ is any of $A$, $B$, $h$, $\Phi$, and $\phi$.  $B(r)$ may be fixed to be anything by a choice of the coordinate $r$, so all the $B_n$ are arbitrary.  All but finitely many of the other coefficients, however,  are determined in terms of the first few.  To be more precise: $h_0=0$ by assumption; $A_0=0$ can be arranged by rescaling $t$ and $\vec{x}$ by a common factor; $h_1=1/L$ can be arranged by rescaling only $t$; $\Phi_0=0$ is a choice one must make in order for $\Phi dt$ to be well-defined at the horizon; and all other coefficients are determined once one chooses $\phi_0$ and $\Phi_1$.  In other words, the solutions to \eno{SecondOrder} and \eno{ZeroEnergy} may be parametrized by $(\phi_0,\Phi_1)$.  It helps our intuition to recall that $\phi_0$ is the value of the scalar field at the horizon, while $\Phi_1$ is essentially the electric field in the radial direction, also evaluated at the horizon.  Given the assumptions just stated, it is easy to show that
 \eqn{QvaluesNear}{
  Q_G &= e^{-B_0} f(\phi_0) \Phi_1  \cr
  Q_N &= {1 \over L} e^{-B_0} \,.
 }
It does not seem to be practical to find solutions of the equations of motion through high-order series expansions, because the expressions for high-order coefficients quickly become quite complicated.  In practice we stopped at fourth order. The resulting expansions are suitable for providing initial values for numerical integration of the differential equations \eno{SecondOrder} at a radius slightly outside the horizon.

\subsection{Far region asymptotics}
\label{FAR}

To discuss asymptotic behavior far from the horizon it helps to pick a gauge, so we fix $B=0$.  We can write the potential $V(\phi)$ as
 \eqn{Vassumptions}{
  V(\phi) = -{12 \over L^2} + {1 \over 2} m_\phi^2 \phi^2 + {\cal O}(\phi^3) \,,
 }
demonstrating $L$ is the radius of curvature of the asymptotic $AdS_5$ geometry, and we can define $\Delta_\phi$, the ultraviolet dimension of the operator dual to $\phi$, according to 
 \eqn{mphi}{
  m_\phi^2 L^2 \equiv \Delta_\phi (\Delta_\phi-4) \,.
 }
 Following \cite{Gubser:2008yx}, we have assumed that this operator is essentially $\tr F^2$ and that the ultraviolet limit is to be matched to QCD at a scale significantly above $T_c$ but not parametrically large.  Thus $\Delta_\phi$ should be only slightly less than $4$, and in our potential $\Delta_\phi \approx 3.93$. 

One can then straightforwardly show that
 \eqn{FarExpansions}{
  A(r) &= \alpha(r) + A^{\rm far}_{2\nu} e^{-2\nu\alpha(r)} + \ldots  \cr
  h(r) &= h^{\rm far}_0 + h^{\rm far}_4 e^{-4\alpha(r)} + 
     h^{\rm far}_{4+2\nu} e^{-(4+2\nu)\alpha(r)} + \ldots  \cr
  \Phi(r) &= \Phi^{\rm far}_0 + \Phi^{\rm far}_2 e^{-2\alpha(r)} + 
    \Phi^{\rm far}_{2+\nu} e^{-(2+\nu)\alpha(r)} + \ldots  \cr
  \phi(r) &= \phi_A e^{-\nu\alpha(r)} (1 + a_\nu e^{-\nu\alpha(r)} + a_{2\nu} e^{-2\nu\alpha(r)}
      + \ldots )  \cr
     &\qquad {}+ \phi_B e^{-\Delta_\phi\alpha(r)} + \ldots
 }
where
 \eqn{alphaDef}{
  \alpha(r) \equiv A^{\rm far}_{-1} {r \over L} + A^{\rm far}_0
 }
and
 \eqn{nuDef}{
  \nu \equiv 4 - \Delta_\phi \,.
 }
The zero-energy constraint \eno{ZeroEnergy} implies
 \eqn{hZeroRule}{
  A^{\rm far}_{-1} = {1 \over \sqrt{h^{\rm far}_0}} \,.
 }
Given \eno{FarExpansions}, it is straightforward to show that the conserved charges in terms of the far-region quantities evaluate to
 \eqn{QvaluesFar}{
  Q_G &= -{2 \over L} A^{\rm far}_{-1} \Phi^{\rm far}_2  \cr
  Q_N &= {2 \over L} A^{\rm far}_{-1} (-2 h^{\rm far}_4 + \Phi^{\rm far}_0
    \Phi^{\rm far}_2) \,.
 }
Thus $A^{\rm far}_{-1}$, $h^{\rm far}_4$, and $\Phi^{\rm far}_2$ can be determined in terms of $h^{\rm far}_0$ and $\Phi^{\rm far}_0$ once $Q_G$ and $Q_N$ are known.  It is notable that in the absence of a scalar (or if for some reason $\phi \to 0$ at the boundary faster than $e^{-2\alpha(r)}$) then the next correction to $h(r)$ after $h^{\rm far}_4 e^{-4\alpha(r)}$ is $h^{\rm far}_6 e^{-6\alpha(r)}$, and the equation of motion for $h$ can be used to show that
 \eqn{hSixRule}{
  h^{\rm far}_6 = {1\over 3} (\Phi^{\rm far}_2)^2 \,.
 }
However, for the solutions we will study numerically, the scalar doesn't vanish fast enough for the $h^{\rm far}_6 e^{-6\alpha(r)}$ term to be interesting.

Evidently, the expansions \eno{FarExpansions} are qualitatively more intricate than the Taylor expansions \eno{Xexpand} around the horizon because the powers of $e^{-\alpha(r)}$ are not (for practical purposes) commensurate, owing to $\nu \approx 0.07$ not being the ratio of small integers.  The fact that $\nu \ll 1$ also leads to some difficulties in finding robust numerical solutions to the equations of motion.  We will discuss these issues at greater length in subsection~\ref{STRATEGY}.

The expansion of $\phi(r)$ in \eno{FarExpansions} is split into the part dual to a deformation (proportional to $\phi_A$) and the part dual to an expectation value (proportional to $\phi_B$).  Each solution carries a series of corrections, which we have shown only for the solution proportional to $\phi_A$.  The correction term $a_\nu e^{-\nu\alpha(r)}$ is present only when $V'''(0) \neq 0$, so for even potentials like \eno{VChoice} the leading correction is $a_{2\nu} e^{-2\nu\alpha(r)}$.  For small enough $\nu$, this correction, and even higher corrections proportional to $\phi_A$, dominate over the $\phi_B e^{-\Delta_\phi \alpha(r)}$ term.  Thus the terms not shown explicitly in the expansion for $\phi(r)$ are subleading either to $\phi_A e^{-\nu\alpha(r)}$ or to $\phi_B e^{-\Delta_\phi \alpha(r)}$---or to both.  In all the other expansions
in \eno{FarExpansions}, the omitted terms are all subleading to the terms shown explicitly.

\subsection{Thermodynamic quantities}
\label{THERMO}

The solutions we are interested in have $\phi_A \neq 0$, because they are to be understood as renormalization group flows triggered by deformation of a very slightly relevant operator.  This is not exactly how QCD works, but it is sufficiently close to be an interesting approximation.  In order to compare solutions meaningfully, one should ideally
arrange for $\phi_A$ always to be the same.  This can be accomplished through a coordinate transformation, provided $\phi(r)$ always has the same sign at the boundary.\footnote{If $\phi(r)$ becomes negative at the boundary, the following expressions for far-zone coefficients and thermodynamic quantities can still be used if $|\phi_A|$ is substituted for $\phi_A$.}  More specifically: suppose we obtain a solution numerically which has some positive value of $\phi_A$.  Then we wish to perform a coordinate transformation on this solution to bring it into the form
 \eqn[c]{GoodFar}{
  d\tilde{s}^2 = e^{2\tilde{A}(\tilde{r})} (-\tilde{h}(\tilde{r}) 
    d\tilde{t}^2 + d\tilde{\vec{x}}^2) + {d\tilde{r}^2 \over \tilde{h}(\tilde{r})}  \cr
  \tilde{A}_\mu d\tilde{x}^\mu = \tilde\Phi(\tilde{r}) d\tilde{t}  \qquad\quad
  \tilde\phi = \tilde\phi(\tilde{r})
 }
where
 \eqn{GoodDemand}{
  \tilde{A}(\tilde{r}) &= {\tilde{r} \over L} + {\cal O}(e^{-2\nu \tilde{r}/L})  \cr
  \tilde{h}(\tilde{r}) &= 1 + \tilde{h}^{\rm far}_4 e^{-4\tilde{r}/L} + 
    {\cal O}(e^{-(4+2\nu) \tilde{r}/L})  \cr
  \tilde\Phi(\tilde{r}) &= \tilde\Phi^{\rm far}_0 + \tilde\Phi^{\rm far}_2 e^{-2\tilde{r}/L} + 
    {\cal O}(e^{-(2+\nu) \tilde{r}/L})  \cr
  \tilde\phi(\tilde{r}) &= e^{-\nu \tilde{r}/L} + {\cal O}(e^{-2\nu \tilde{r}/L})
 }
Setting $ds^2 = d\tilde{s}^2$, $A_\mu dx^\mu = \tilde{A}_\mu d\tilde{x}^\mu$, and $\phi(r) = \tilde\phi(\tilde{r})$, one finds immediately that
 \eqn{GoodCoords}{
  \tilde{t} &= \phi_A^{1/\nu} \sqrt{h^{\rm far}_0} \, t  \cr
  \tilde{\vec{x}} &= \phi_A^{1/\nu} \vec{x}  \cr
  {\tilde{r} \over L} &=  \alpha(r) - {\log(\phi_A^{1/\nu})} =  A^{\rm far}_{-1} {r \over L} + A^{\rm far}_0 - 
    {\log(\phi_A^{1/\nu})}
 }
and
\eqn{GoodFuncs}{
  \tilde{A}(\tilde{r}) &=A(r) - {\log(\phi_A^{1/\nu})}   \cr
  \tilde{h}(\tilde{r}) &= {1 \over h_0^{\rm far} }\, h(r)  \cr
  \tilde\Phi(\tilde{r}) &= {1 \over \phi_A^{1/\nu}  \sqrt{h_0^{\rm far}}} \, \Phi(r) \,,
 }
which implies
 \eqn{GoodCoefs}{
  \tilde\Phi_0^{\rm far} &= {\Phi^{\rm far}_0 \over \phi_A^{1/\nu} \sqrt{h^{\rm far}_0}}  \cr
  \tilde\Phi_2^{\rm far} &= {\Phi^{\rm far}_2 \over \phi_A^{3/\nu} \sqrt{h^{\rm far}_0}}  \cr
  \tilde{h}_4^{\rm far} &= {h^{\rm far}_4 \over \phi_A^{4/\nu} h^{\rm far}_0} \,.
 }
Intensive thermodynamic quantities can now be readily extracted using these relations along with 
 standard expressions in the $(\tilde{t},\tilde{\vec{x}},\tilde{r})$ coordinate system and the assumptions stated following \eno{Xexpand}:
 \eqn{GoodIntensive}{
  T &= {e^{\tilde{A}(\tilde{r}_H)} \over 4\pi} 
    \left( {d\tilde{h} \over d\tilde{r}} \right)_{\tilde{r}=\tilde{r}_H} = 
     {1 \over 4\pi} {1 \over L \phi_A^{1/\nu} \sqrt{h^{\rm far}_0}}  \cr
  \mu &= {\tilde\Phi^{\rm far}_0 \over L} = 
    {\Phi^{\rm far}_0 \over L \phi_A^{1/\nu} \sqrt{h^{\rm far}_0}} \,.
 }
Densities of extensive thermodynamic quantities can likewise be computed:
 \eqn{GoodExtensive}{
  s &= {2\pi \over \kappa^2} e^{3\tilde{A}(\tilde{r}_H)} = 
    {2\pi \over \kappa^2} {1 \over \phi_A^{3/\nu}}  \cr
  \rho &= -{ \tilde\Phi^{\rm far}_2 \over \kappa^2} = -{ \Phi^{\rm far}_2 \over 
    \kappa^2 \phi_A^{3/\nu} \sqrt{h^{\rm far}_0}}  \,.
 }
 Thus knowledge of the four asymptotic scaling parameters $\phi_A$, $h_0^{\rm far}$, $\Phi_0^{\rm far}$ and $\Phi_2^{\rm far}$ determines the standard thermodynamic variables $T, \mu, s$ and $\rho$.   Subleading parameters in the field expansions will depend on these in general.  For example, using the constancy of the Noether charge \eno{NoetherConstraint} and its asymptotic expressions \eno{QvaluesNear} and \eno{QvaluesFar}, for $h_4^{\rm far}$ one can show 
 \eqn{EnergyPressure}{
{h^{\rm far}_4 \over \phi_A^{4/\nu} h^{\rm far}_0} = \tilde{h}_4^{\rm far} = - {\kappa^2 L \over 2} (sT + \mu \rho) = - {\kappa^2 L \over 2} (\epsilon + p) \,,
 }
 where in the last step we used the thermodynamic relation (\ref{ThermRel}) to express the result in terms of the sum of the pressure and energy density.
 Note that this is the only combination of $\epsilon$ and $p$ we have access to from these calculations.  The pressure by itself is equivalent to the free energy density (\ref{PressureFreeEnergy}), which to calculate we would need to evaluate the full renormalized action including counterterms to cancel divergences.  It is possible to get the results we're interested in --- in particular the position of the critical point and the values of its critical exponents --- with just $T$, $\mu$, $s$, and $\rho$.

We can also use the Gauss charge to relate a certain combination of the near-horizon parameters $(\phi_0, \Phi_1)$ to the asymptotic parameters, and thus the thermodynamics.  One finds that the Gauss charge is proportional to the inverse of the entropy per baryon:
\eqn{EntropyBaryon}{
Q_G = f(\phi_0) \Phi_1 = {4 \pi \over L } {\rho \over s} \,.
}
This is the only analytic relation between the initial conditions at the horizon and the thermodynamic parameters.

\subsection{Numerical strategy}
\label{STRATEGY}

It is straightforward in principle to obtain a numerical black hole solution by integrating the second-order equations of motion \eno{SecondOrder} starting at some point slightly outside the horizon with the functions and their derivatives initialized from the horizon series expansions described in section~\ref{NEAR},
with initial conditions $(\phi_0, \Phi_1)$.
Then a fit can be performed of the numerically known functions $A(r)$, $h(r)$, $\Phi(r)$, and $\phi(r)$ to the asymptotic forms \eno{FarExpansions} to extract the quantities $h^{\rm far}_0$, $\Phi^{\rm far}_0$, $\Phi^{\rm far}_2$, and $\phi_A$ in terms of which $T$, $\mu$, $s$, and $\rho$ can be determined.  Thus for each input value of $(\phi_0, \Phi_1)$, we obtain a black hole characterized by thermodynamic quantities $(T, \mu, s, \rho)$.  Certain values of $(\phi_0, \Phi_1)$ may lead to a solution that does not converge to an asymptotically-$AdS_5$ solution.  Typically, these spacetimes are singular.  They are not of the class we are interested in, so they are discarded.

Numerical integrations can be made vastly more efficient by noting that $h(r)$ and $\Phi(r)$ converge much faster to their asymptotic values than $\phi(r)$ and $A'(r)$.  A good strategy, then, is to figure out the value $r=r_*$ beyond which the non-constant corrections to $h(r)$ and $\Phi(r)$ have no more influence on the equations of motion for $A$ and $\phi$ than round-off errors do; then join a solution of the full equations of motion from a point just outside the horizon to $r=r_*$ to a solution to simplified equations of motion, obtained by replacing $h$ by $h^{\rm far}_0$ and $\Phi$ by $\Phi^{\rm far}_0$, from $r=r_*$ to a value of $r$ large enough to reliably compute $\phi_A$.  As discussed following \eno{QvaluesFar}, $\Phi^{\rm far}_2$ can be determined once $h^{\rm far}_0$ and $\Phi^{\rm far}_0$ are known.  Thus in order to extract $T$, $\mu$, $s$, and $\rho$ using \eno{GoodIntensive}-\eno{GoodExtensive}, the only quantities one needs from numerics are $h^{\rm far}_0$, $\Phi^{\rm far}_0$, and $\phi_A$.  We implemented the strategy described here in Mathematica, where the basic ODE's \eno{SecondOrder} are solved using {\tt NDSolve}.

\section{Quark susceptibility at zero chemical potential}
\label{SUSCEPTIBILITY}

Because lattice calculations at finite chemical potential are problematic, it has been difficult to make precise predictions for the behavior of QCD off the $T$-axis.  However, at $\mu =0$, lattice studies have been carried out extensively.  The potential $V(\phi)$ from \cite{Gubser:2008ny,Gubser:2008yx} was engineered to reproduce the equation of state $s(T)$ known from lattice simulations.

We would like to also constrain the gauge kinetic function $f(\phi)$ using known lattice results at $\mu=0$.   The extrapolation to finite $\mu$ is then completely determined by known physics at $\mu=0$, and represents the unique prediction for the phase diagram of the large-$N$ gauge theory defined to emulate the thermodynamics of QCD on the $T$-axis. 

The gravity calculation of $s(T)$  at $\mu =0$ is completely insensitive to $f(\phi)$, since the gauge field is zero in these solutions.  Instead we may examine the quark susceptibility \eno{ChiEqn} at vanishing $\mu$ as a function of temperature, as this has also been calculated extensively on the lattice and as we will see, depends on the choice of $f(\phi)$.  In section~\ref{FORMULA} we find a gravity formula for the quark susceptibility at zero chemical potential, and in section~\ref{TAXIS} we use this to justify our choice of $f(\phi)$.

\subsection{A formula for quark susceptibility}
\label{FORMULA}

The black holes with $\mu = 0$ have vanishing gauge field $A_\mu$, and $\rho = 0$ as well.  To calculate the quark susceptibility \eno{ChiEqn}, we make use of the key observation that the gauge field equation of motion is linear and homogeneous in $\Phi$, while $\Phi$ appears only quadratically in the remaining equations \eno{SecondOrder}.  
We thus proceed by treating $\Phi$ as a linear perturbation, solving the gauge field equation in the fixed background of the $\mu = 0$ black hole, and then
determine $\chi_2$ by noting that on the $T$-axis, its definition \eno{ChiEqn} becomes
 \eqn{ChiSimpler}{
  \chi_2(\mu \!=\! 0) = \lim_{\mu \to 0} {\rho(\mu) \over \mu} \,.
 }
Moreover, in the linearized approximation, the overall normalization of $\Phi$ is arbitrary as far as the equations of motion are concerned and will cancel out of \eno{ChiSimpler}, so we can set it to $\Phi_1 = 1/L$. 

We can in fact obtain a formula for \eno{ChiSimpler} that reduces to quantities only involving the metric and scalar, which are unchanged in the linearized approximation and thus can be taken from the solution for the 
background $\mu=0$ black hole.
Since it is common in the literature to plot $\chi_2$ normalized by $T^2$, which approaches a constant at large $T$, we will find a formula for $\hat\chi_2 \equiv \chi_2/T^2$.
Using equations  \eno{GoodIntensive}-\eno{GoodExtensive}, we have
 \eqn{ChiOverT2}{
 \hat\chi_2(\mu  \!=\!  0)  = { \rho \over \mu T^2} = 
       - {(4 \pi)^2 L^3 \over \kappa^2} { \Phi_2^{\rm far} h_0^{\rm far} \over \Phi_0^{\rm far}} \,,
 }
 and making use of the expression \eno{QvaluesFar} for the Gauss charge $Q_G$, we may simplify \eno{ChiOverT2} to
 \eqn{ChiOverT2Gauss}{
 { \hat\chi_2}(\mu  \!=\!  0) = {8 \pi^2 L^4 \over \kappa^2} (h^{\rm far}_0)^{3/2} {Q_G \over \Phi^{\rm far}_0} \,.
 }
Now recall that $\Phi \to 0$ at the horizon $r=r_H$.  As a result,
\eqn{QGRatio}{
\Phi^{\rm far}_0 = \int^{\infty}_{r_H} dr \,  \Phi' = Q_G \int_{r_H}^\infty dr \, e^{-2A} f(\phi)^{-1} \,, 
}
where in the second step we have employed the definition \eno{GaussConstraint} of the Gauss charge.  Plugging \eno{QGRatio} into \eno{ChiOverT2Gauss} results in
\eqn{Chi2Final}{
\hat\chi_2(\mu  \!=\!  0) = {8 \pi^2 L^4 \over \kappa^2} {(h^{\rm far}_0)^{3/2} \over \int_{r_H}^\infty dr \, e^{-2A} f(\phi)^{-1}} \,.
}
Note at this point that all explicit dependence on the gauge field $\Phi$ has dropped out.  One can illuminate this further by noting that
\eqn{sT3}{
{s \over T^3} = {128 \pi^4 L^3 \over \kappa^2} \, (h^{\rm far}_0)^{3/2} \,,
}
so that finally
\eqn{chi2sT3}{
\hat\chi_2(\mu  \!=\!  0) = {L \over 16 \pi^2} \, {s \over T^3} \, {1 \over \int_{r_H}^\infty dr \, e^{-2A} f(\phi)^{-1}}  \,.
}
This expression may now be evaluated on $\mu=0$ black holes directly, without having to solve the linearized $\Phi$ equation explicitly.

Our final expression \eno{chi2sT3} is suggestive because in lattice simulations, $s/T^3$ and $\hat\chi_2 \equiv \chi_2 / T^2$ have qualitatively similar behavior as functions of temperature at $\mu=0$: both start near zero for low temperatures, then rapidly cross over to a large value in the region of $T_c$, and asymptote to a finite value at large $T$.  Hence from (\ref{chi2sT3}) we come to expect the realistic behavior of $\hat\chi_2$ will to some extent be inherited from the analogous behavior of $s/T^3$.  

The effects of the integral in the denominator of \eno{chi2sT3} do play an important role, however.  This integral introduces a dependence of the quark susceptibility on the function $f(\phi)$, which $s/T^3$ alone was insensitive to.  Thus differences between the functional forms of $\hat\chi_2$ and $s/T^3$ are due entirely to the effects of $f(\phi)$.

\subsection{Matching to lattice data at zero chemical potential}
\label{TAXIS}

\begin{figure}
\begin{center}
\includegraphics[scale=0.5]{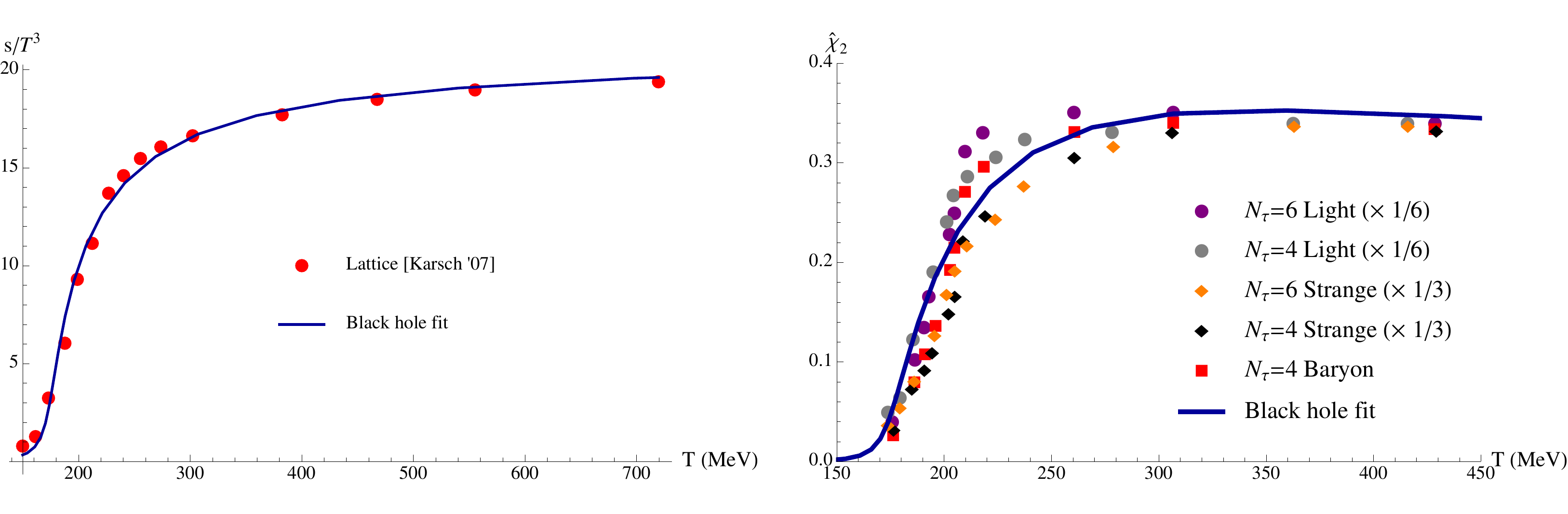}
\caption{The normalized entropy $s/T^3$ and quark susceptibility $\hat\chi_2 \equiv \chi_2/T^2$ at $\mu=0$, computed on the lattice and fit by black holes in the gravity theory defined by our choices of $V(\phi)$  and $f(\phi)$ (equations \eno{VChoice} and \eno{fChoice}).  Lattice data is taken from \cite{Karsch:2007dp}.
\label{ZeroMuComparison}}
\end{center}
\end{figure}

We now discuss the matching of lattice data to black hole results at $\mu=0$ and justify our choice \eno{fChoice} for $f(\phi)$.  Since the precise field theory dual of our model is unknown, we will not try to translate the quantities $\kappa$, $L$ into field theory language.  Instead, we will make the arbitrary choice $\kappa = L = 1$ and parametrize our ignorance by allowing separate overall constant rescalings between the lattice quantities and the black hole quantities:
 \eqn{sTrescale}{
  [s]_{\rm lattice} = \lambda_s [s]_{\rm BH} \,, \quad
  [T]_{\rm lattice} = \lambda_T [T]_{\rm BH} \,, \quad
  [\rho]_{\rm lattice} = \lambda_\rho [\rho]_{\rm BH} \,, \quad
  [\mu]_{\rm lattice} = \lambda_\mu [\mu]_{\rm BH} \,.
 }
As described previously, to compute the entropy density at $\mu=0$ one doesn't need any information about $f(\phi)$ at all.  In figure~\ref{ZeroMuComparison}A we show how the entropy density compares between lattice and black holes based on the potential \eno{VChoice}.  For lattice data we used the right hand plot in Figure~3 of \cite{Karsch:2007dp}\footnote{We chose \cite{Karsch:2007dp} to have a definite lattice result to compare to, but there is still disagreement in the literature; for another determination of the equation of state, see \cite{Borsanyi:2010cj}.  We expect small changes to our model could accomodate variations in the lattice results.}, with points from asqtad $N_\tau=6$ simulations from $T=150\,{\rm MeV}$ out to $T=382\,{\rm MeV}$, and then points from p4 $N_\tau=6$ simulations out to $T=720\,{\rm MeV}$.  We determined $T_c \approx 191\,{\rm MeV}$ as the temperature at which $s/T^3$ reaches $1/e$ of its largest value as obtained from the highest temperature data point.\footnote{$T_c \approx 191\,{\rm MeV}$ is somewhat larger than the value $T_c \approx 175\,{\rm MeV}$ mentioned in section~\ref{INTRODUCTION}; however it is in line with estimates of \cite{Cheng:2006qk}.  Lower values for $T_c$ are favored, for example, in \cite{Aoki:2006br,Aoki:2009sc,Borsanyi:2010zi}.  We do not aim here to probe the apparent discrepancy; instead we are largely opting for the higher values because all the lattice data we use directly is from \cite{Karsch:2007dp}.}  For black hole data, we constructed black holes starting with our standard choice \eno{VChoice} of scalar potential, and for $\phi_0$ ranging from $1.5$ to $7.5$ in $20$ steps, uniform on a log scale. 

 Our conventions are for all lattice quantities to have units which are powers of ${\rm MeV}$, while with $\kappa=L=1$, all black hole quantities are dimensionless.  Thus $\lambda_s$ and $\lambda_T$ have units which are also powers of ${\rm MeV}$.  We found a good fit between lattice data and black holes with
 \eqn{lambdaTsValues}{
  \lambda_s = \left( 121\ {\rm MeV} \right)^3 \,, \qquad
  \lambda_T = 252 \  {\rm MeV} \,.
 }
Turning to the quark susceptibility, in figure~\ref{ZeroMuComparison}B we show how susceptibilities computed starting from the choice \eno{fChoice} for $f(\phi)$ compare with lattice data.  For lattice data we used the same reference as for the entropy  \cite{Karsch:2007dp}, with light quark results from the left hand plot in Figure~5, and strange quark and total baryon number results from the two sides of Figure~6; we scaled the light quark and strange quark curves appropriately to asymptote to the same value as that of the baryon number, $\hat\chi_2 = 1/3$, at high temperatures.
 For black hole data we used the same black holes as in the entropy plot, and we employed \eno{chi2sT3} with $L=1$ (and implicitly also $\kappa=1$ as before).  We used the value of $\lambda_T$ in \eno{lambdaTsValues} to rescale the temperature axis, and we adjusted the overall scale of $\hat\chi_2$ arbitrarily to optimize the fit to lattice over the range shown in the figure~\ref{ZeroMuComparison}B.  
 The rescaling of the susceptibility is thus
 \eqn{RescaleChi}{
  [\chi_2]_{\rm BH} \equiv \left[ {\partial\rho \over \partial\mu} \right]_{\rm BH} = 
    {\lambda_\mu \over \lambda_\rho} [\chi_2]_{\rm lattice} \,.
 }
Thus, knowing $[\chi_2]_{\rm BH}$ and $[\chi_2]_{\rm lattice}$ at the same temperature tells us $\lambda_\mu / \lambda_\rho$.  In order to find $\lambda_\mu$ and $\lambda_\rho$ separately, we must recall that the relation for the free energy \eno{FreeEnergy}
holds equally in lattice units and in the black hole setup.  Thus
 \eqn{lambdaRelations}{
  \lambda_T \lambda_s = \lambda_\mu \lambda_\rho = \lambda_\epsilon = \lambda_f \,.
 }
Putting \eno{RescaleChi} and \eno{lambdaRelations} together, we find
 \eqn{RescaleChiHat}{
  [\hat\chi_2]_{\rm BH} \equiv \left[ {1 \over T^2} {\partial\rho \over \partial\mu} \right]
    = {1 \over \lambda_T^2} { \lambda_\mu^2    \over     \lambda_\rho \lambda_\mu}
      [\hat\chi_2]_{\rm lattice}
    = { \lambda_T \lambda_\mu^2 \over \lambda_s } [\hat\chi_2]_{\rm lattice} \,,
 }
which can be recast as
 \eqn{FoundLambdaMu}{
  \lambda_\mu = \sqrt{{\lambda_s \over \lambda_T} {[\hat\chi_2]_{\rm BH} \over 
    [\hat\chi_2]_{\rm lattice}}} \,.
 }
Let us now describe how we arrived at the choice \eno{fChoice} for functional form for the gauge kinetic function.  The value of $f(\phi)$ near the horizon is particularly important, because the factor $e^{-2A}$ in the integral \eno{chi2sT3} puts significant weight on the near-horizon region.  
In particular, if $f(\phi)$ is large at the horizon, the integral will be relatively small compared to when $f(\phi)$ is small at the horizon.  Since $\hat\chi_2$ stays close to its high-temperature value down to a lower temperature than $s/T^3$ before plunging rapidly to small values, a reasonable conjecture is that as $\phi$ goes from $0$ to positive values, one needs $f(\phi)$ first to increase as a function of $\phi$, then to decrease rapidly.  The functional form \eno{fChoice} was chosen with these desired features in mind, and also with the thought that asymptotically exponential behavior at large $\phi$ is typical of supergravity theories.

Operationally, the way we determined $\lambda_\mu$ was to use the correct Stefan-Boltzmann value $\hat\chi_2 = 1/3$ for baryon number as the value for $[\hat\chi_2]_{\rm lattice}$, and to evaluate $[\hat\chi_2]_{\rm BH}$ at $T = 460\,{\rm MeV}$.  This is a reasonable approach because the lattice data converges quickly to the Stefan-Boltzmann value at high temperature.  The result is
 \eqn{LambdaMuRhoValues}{
  \lambda_\mu = 972\ {\rm MeV} \,, \qquad
  \lambda_\rho = \left( 77\ {\rm MeV} \right)^3 \,.
 }
Again it should be emphasized that our choice \eno{fChoice} of $f(\phi)$ is to a degree {\it ad hoc}, and it should be understood as providing a proof of principle that an approximate fit to $\chi_2$ can go with a critical endpoint in the $T$-$\mu$ plane based on AdS/CFT techniques.

\section{Searching for the critical point}
\label{SEARCH}

Having settled on a functional form for $V(\phi)$ and $f(\phi)$ by matching to lattice thermodynamics at $\mu=0$, our Lagrangian is now completely determined.  We can next turn to numerically solving for a set of black holes to fill in the phase diagram.  An expectation is that the crossover that takes place on the $T$-axis is sharpened into a first-order line lying out in the $T$-$\mu$ plane, and that this first-order line ends at a critical point somewhere in the vicinity of the crossover.  Our first task therefore is to search for this critical point.

\subsection{Scanning the thermodynamics of black holes}

For a first pass at mapping out the thermodynamic behavior of black holes across the $T$-$\mu$ plane, we generated approximately $2500$ numeric solutions to the equations \eno{SecondOrder} and \eno{ZeroEnergy}, seeded by initial conditions near the horizon as described in section~\ref{NEAR}.  Each solution is specified by the value of $(\phi_0,\Phi_1)$ that was used to generate the near-horizon asymptotics.  We remind the reader that $\phi_0$ is the value of the scalar field at the horizon, which we took always to be positive, and $\Phi_1$ is essentially the electric field at the horizon pointing upward in the fifth dimension.  We worked exclusively in the gauge $B=0$.

To choose a suitable range for $\phi_0$, we note first of all that the fits discussed in the previous section involved values of $\phi_0$ no larger than $7.5$.  We went from $\phi_0 = 1$ to $\phi_0 = 15$ in order to obtain the best global picture of the thermodynamics that we could, but any features seen at $\phi_0$ significantly larger than $7.5$ should be regarded with some degree of skepticism, since in principle one could adjust $V(\phi)$ and/or $f(\phi)$ for $\phi>7.5$ to make any desired phenomenon occur in that region.\footnote{In fact, constraints on $f(\phi)$ come from a narrower range of $\phi$, extending only up to $\phi_0=5$. Thus, baryon-specific physics is most reliably studied in our model at values of $\phi_0$ no greater than $5$.}  It is notable, however, that both $V(\phi)$ and $f(\phi)$ are fairly featureless for $\phi \gsim 4$, both being close to a simple exponential function over that domain.

To choose a suitable range for $\Phi_1$, we demonstrate that there is an upper bound on possible $\Phi_1$ values leading to an asymptotically-AdS black hole.
To see this, first note that the first equation of \eno{SecondOrder} with $B=0$ shows that $A$ is concave down as a function of $r$.  But it must be increasing at large $r$ in order for the spacetime to be asymptotically $AdS_5$.  Therefore $A$ must be increasing at the horizon, which is to say  $A_1 > 0$.   
Using the zero-energy constraint \eno{ZeroEnergy}, $A_1$ can be re-expressed as
 \eqn{AoneExpression}{
  A_1 = -{L \over 6} \left[ 2
  V(\phi_0) + f(\phi_0) \Phi_1^2 \right] \,.
 }
Because $V(\phi_0) < 0$ and $f(\phi_0) > 0$, this puts an upper bound on $\Phi_1$:
 \eqn{PhiOneMax}{
  |\Phi_1| < \Phi_{1,{\rm max}} \equiv \sqrt{-{2V(\phi_0) \over f(\phi_0)}} \,.
 }
In practice we scanned black hole solutions from $\Phi_1$ just slightly greater than $0$ up to $0.9 \Phi_{1,\rm max}$.

\begin{figure}
  \centerline{\includegraphics[width=7in]{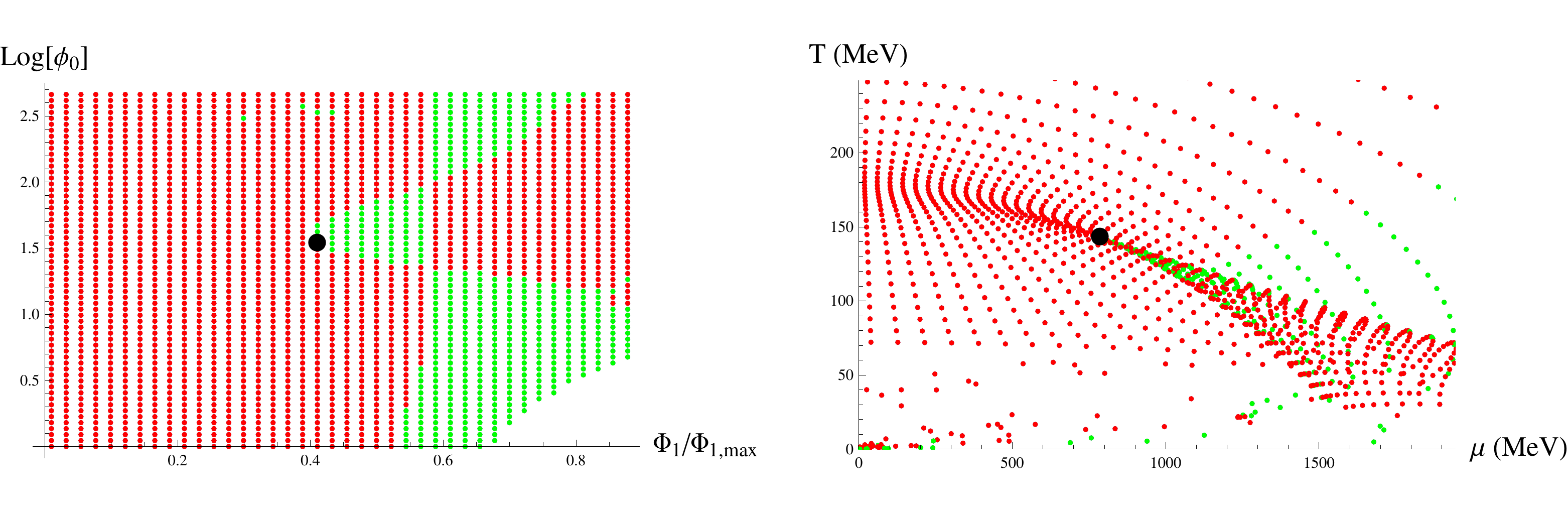}}\begin{picture}(0,0)(0,0)\put(90,0){\Large (A)}\put(350,0){\Large (B)}\end{picture}
  \caption{Numerically generated black holes.  Each dot represents a numerically generated solution.  If the Jacobian $J$ defined in \eno{Jdef} is positive for this solution, then the dot is red.  If $J<0$, it is green.  The bold black circle is the critical endpoint.}\label{GridPlots}
 \end{figure}
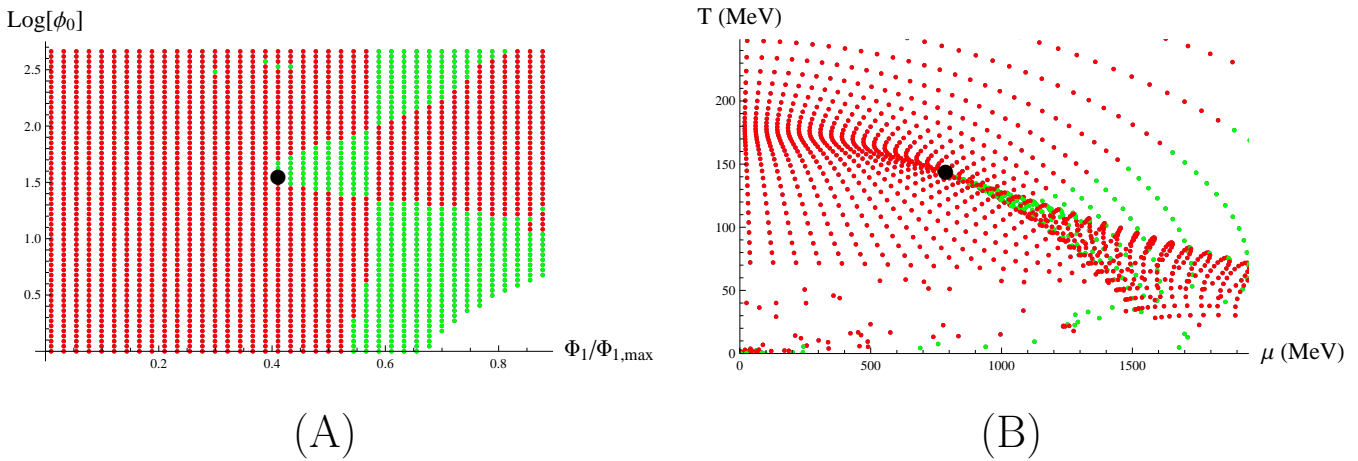

Figure~\ref{GridPlots} shows the results of our numerical scan of the $T$-$\mu$ plane. We examined $61$ values of $\phi_0$ between $1$ and $15$, uniformly spaced on a log scale.  For each value of $\phi_0$ so obtained, we examined $41$ evenly spaced values of $\Phi_1/\Phi_{1,{\rm max}}$.  A small fraction of the values so chosen failed to produce good black hole solutions, 
generally because $A$ failed to be monotonically increasing, and are simply omitted from the plots.

\subsection{Locating the critical point}
\label{LOCATING}

To locate the critical point, we must think a little about what we expect to find in the vicinity of the first-order line.  When there are competing phases in a thermodynamic system, only the one minimizing the free energy is the true ground state.  However, there is no reason to think our black hole solution-generating method will discover only the true ground state solutions.  Due to chiral symmetry not being exact and the presence of the crossover, there is no invariant distinction between the two sides of the first-order line that could correspond to a difference in topology or other invariant distinction between the phases on the gravity side; the distinct phases will be continuously connected in the space of solutions.  Since we are just solving the equations of motion, we expect to find all extrema of the free energy.  

In general, extrema of the free energy include not just locally stable minima, but also any thermodynamically unstable saddle points or maxima.  Far from the first-order line on the $T$-$\mu$ plane we expect only one solution to the equations of motion; 
in the vicinity of the first-order line, however,
we expect to find $\rho$ and $s$ to be multivalued.  Note that this will be true not only on top of the first-order line, but also merely near it, as the free-energetically-unfavored phase will persist for some distance on the phase diagram before ceasing to exist as a solution.  The first-order line ends precisely at the $(T_c, \mu_c)$ where this multivalued behavior ceases; this is the critical point.

\begin{figure}
\begin{center}
\includegraphics[scale=0.6]{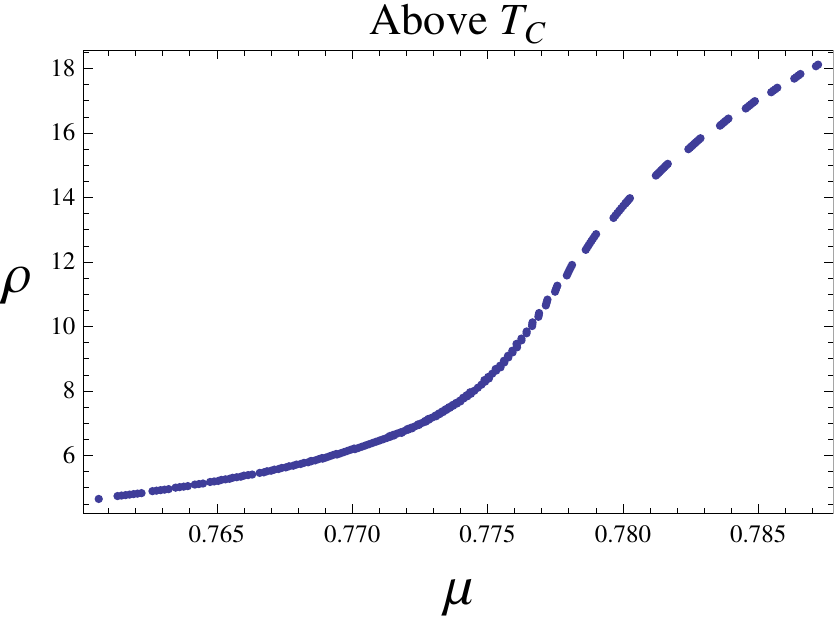}
\includegraphics[scale=0.5]{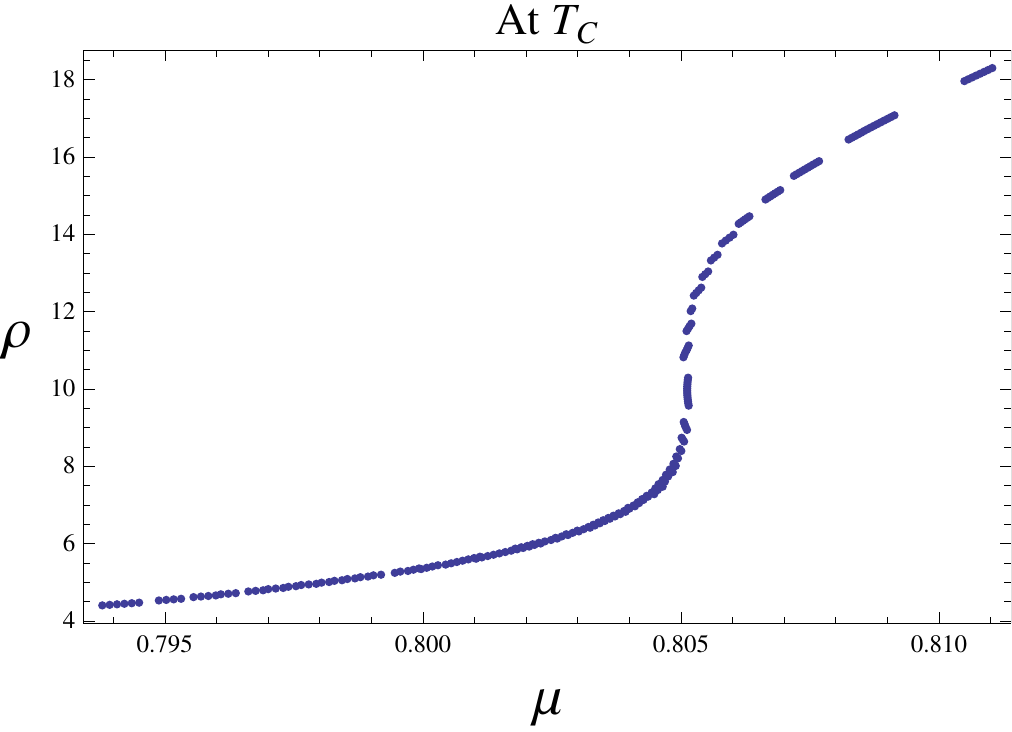}
\includegraphics[scale=0.5]{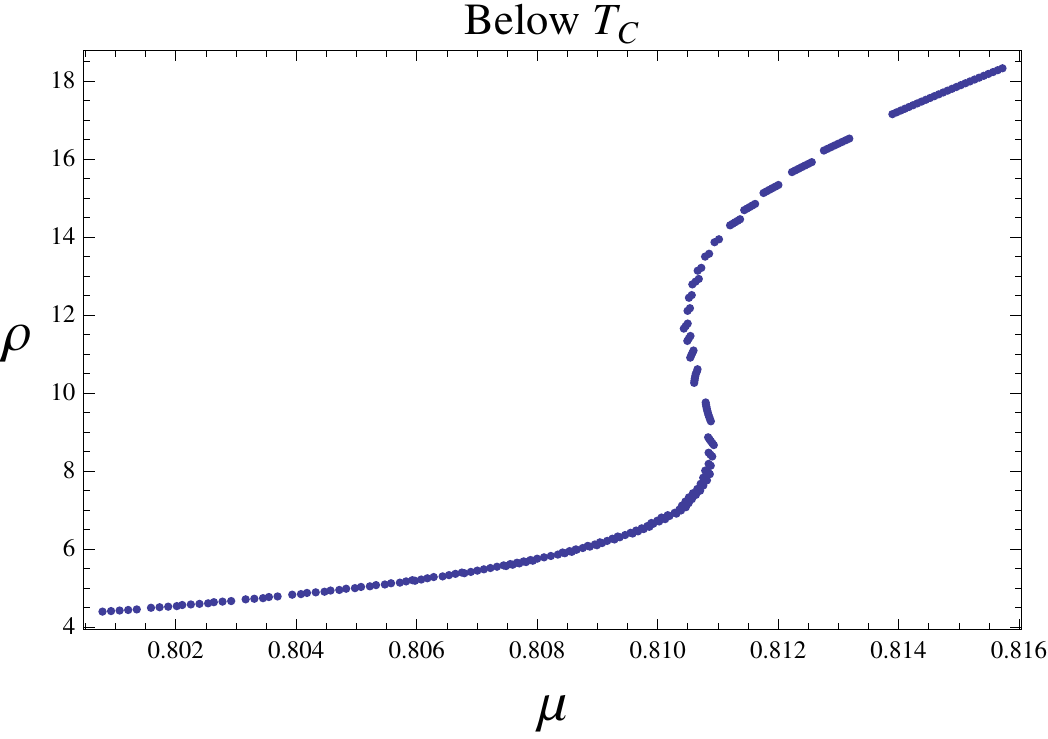}
\caption{The baryon density $\rho$ as a function of chemical potential $\mu$ for several values of $T$ near the critical point.  For $T > T_c$, $\rho(\mu)$ is single-valued (left), while for $T < T_c$ it is multi-valued (right).  At $T= T_c$ the slope is infinite (middle).
\label{fig:rhoMu}}
\end{center}
\end{figure}

Thus if we consider a constant-$T$ slice of the phase diagram with $T > T_c$ and vary $\mu$, this isotherm will miss the first-order line and the functions $\rho(\mu)$ and $s(\mu)$ will be single-valued (although for $T$ close to $T_c$ they will display crossover-type behavior).  But for $T < T_c$, the isotherm will intersect the first order line and we expect $\rho(\mu)$ and $s(\mu)$ to be multivalued near $\mu_c$.  The simplest behavior that still increases at both large and small $\mu$ is an  ``S"-shape, and this is what we observe; see figure~\ref{fig:rhoMu}.  For such behavior there are three solutions at a given $\mu$.    Since the slope of the curve is just the quark susceptibility \eno{ChiEqn}, we see that two of the solutions have $\chi_2 > 0$ and thus may be thermodynamically stable \eno{LocalStab}; these are the candidate phases.  The middle solutions, however, have $\chi_2 < 0$ and must be thermodynamically unstable.\footnote{According to the correlated stability conjecture (CSC) \cite{Gubser:2000ec,Gubser:2000mm}, such black hole solutions will also have dynamical instabilities, corresponding to the black hole gaining total entropy by locally redistributing charge and energy subject to global conservation of these quantities.  In the black hole literature this is known as the Gregory-Laflamme instability \cite{Gregory:1993vy,Gregory:1994bj}.}
Precisely for $T = T_c$ the curve will cease to be multivalued, as the three solutions coalesce into one; the curve $\rho(\mu)$ will have an infinite slope, indicating a divergence in the quark susceptibility at the critical point.

We will locate the critical point by looking for the thermodynamically {\em unstable} solutions that characterize the vicinity of the first-order line.  We can identify the unstable solutions by calculating the Jacobian of the susceptibility matrix \eno{SusceptMatrix}:
 \eqn{Jdef}{ 
  J \equiv \det {\cal S} =  \partial(s,\rho)/\partial(T,\mu) \,.
 }
For a thermodynamically stable black hole, equation \eno{LocalStab} is satisfied and the Jacobian \eno{JacSuscept} is manifestly positive.  If it flips sign to $J < 0$, we have necessarily found a thermodynamically unstable branch.  
Once we find the thermodynamically unstable black holes, we look to see whether they map to a narrow line-like region on the $T$-$\mu$ plane; the critical point is then the values $(T_c, \mu_c)$ where this line ends.  We should also be able to see the two stable phases mapping to the same locus on the phase diagram from elsewhere in $(\phi_0, \Phi_1)$.

We can calculate the Jacobian $J$ by finite differences.
Since the black holes were scanned on a rectangular grid, we can label them with indices $ij$, where $i$ determines the value of $\phi_0$ and $j$ determines the value of $\Phi_1/\Phi_{1,{\rm max}}$.  In order to compute $J$ for the black hole labeled $ij$, we first computed
 \eqn{TwoJs}{
  J^{T\mu}_{ij} &\equiv \det\begin{pmatrix} T_{i+1,j}-T_{i,j} & T_{i,j+1}-T_{i,j} \\
    \mu_{i+1,j}-\mu_{i,j} & \mu_{i,j+1}-\mu_{i,j} \end{pmatrix}  \cr\noalign{\vskip2\jot}
  J^{s\rho}_{ij} &\equiv \det\begin{pmatrix} s_{i+1,j}-s_{i,j} & s_{i,j+1}-s_{i,j} \\
    \rho_{i+1,j}-\rho_{i,j} & \rho_{i,j+1}-\rho_{i,j} \end{pmatrix} \,.
 }
Then $J^{s\rho}_{ij} / J^{T\mu}_{ij}$ is the finite difference approximation to the Jacobian $J$ in \eno{Jdef}.  The results are shown in figure~\ref{GridPlots}.

It is clear from figure~\ref{GridPlots}A that there is a region of unstable black holes stretching down to $\Phi_1 \approx 0.4 \Phi_{1,{\rm max}}$.  It is this region which we are most interested in, because when mapped to the $T$-$\mu$ plane it becomes a narrow region that ends in a cusp.
Moreover, we do indeed find two other sets of black holes with $J > 0$ mapped to the same locus, and hence we identify it as the first-order line.  The point of this cusp is then the critical endpoint, which we show as a bold black circle.  It occurs at the values
 \eqn{CritValues}{
  (T_c,\mu_c) \approx (143\,{\rm MeV},783\,{\rm MeV}) \,.
 }
We have used the multipliers $\lambda_T$ and $\lambda_\mu$ from \eno{lambdaTsValues} and \eno{LambdaMuRhoValues} to express $T$ and $\mu$ in units of ${\rm MeV}$.  Points at the critical point come from initial conditions in the vicinity of
\eqn{CritInitialConds}{
(\phi_0, \Phi_1/\Phi_{1,{\rm max}}) \approx (4.84, 0.40) \,.
}
Note that the value for $\phi_0$ is within the range probed by the $\mu=0$ solutions described in section~\ref{TAXIS}---though not by much, if one goes by the values of $\phi$ over which $f(\phi)$ is meaningfully constrained by lattice data.

In summary, we have identified a candidate critical point and first-order line.  As we study it in more detail in the next section, examining the behavior of densities and susceptibilities, this identification will be amply confirmed.

Before moving on, let us consider the other thermodynamically unstable black holes found in our scan.   The unstable black holes described so far are associated with a failure of the map $(\phi_0,\Phi_1) \to (T,\mu)$ to be invertible, which is to say a sign change in $J^{T\mu}_{ij}$.   Only with such multiple covering can you jump abruptly from one solution to another at the same $(T,\mu)$ but different $(s,\rho)$: the {\it sine qua non} of first-order phase transitions.  As one proceeds further to the right in the $T$-$\mu$ plane, one encounters a broader region of unstable black holes immediately above the multiply-covered region. This region is unstable due to a change of sign in $J^{s\rho}_{ij}$, meaning it is the map of $(\phi_0,\Phi_1) \to (s,\rho)$ that is not invertible.  This sign change causes black hole instabilities, presumably of the Gregory-Laflamme type \cite{Gregory:1993vy,Gregory:1994bj}, but no first-order line. 
Correspondingly, there are no stable black holes in this region of the phase diagram.\footnote{It is interesting to note that $J^{s\rho}_{ij}$ tends to change sign close to $\Phi_1/\Phi_{1,{\rm max}} \approx 0.6$ for a fairly wide range of other choices for $f(\phi)$ that we examined numerically.}

The absence of stable black holes in our model at large $\mu$ (roughly, larger than $\mu = 1100\,{\rm MeV}$) and $T$ not too big is actually a good thing.  It is in approximately this region that one might reasonably expect color superconductivity and/or related phenomena to set in: see for example Figure~7 of \cite{Stephanov:2004wx} and Figure~1 of \cite{Aggarwal:2010cw}.  Black holes based on the lagrangian \eno{LwithF} are not likely to capture such phenomena.  However, it is comforting to note that in cases where black hole superconductivity is understood (where the condensate breaks a $U(1)$ gauge symmetry in the bulk), the superconducting instability competes against Gregory-Laflamme instabilities, and one generally must pass beyond minimal supergravity lagrangians to see the superconducting instabilities: see for example \cite{GubserRome,Gubser:2009qm}.  Thus all our findings are at least qualitatively consistent with consensus expectations for the QCD phase diagram.

\section{Analysis of the Critical Point}
\label{ANALYSIS}

Having found the critical point, our final task is to determine its critical exponents.  To achieve this, we first construct a large data set which densely populates the critical region. This collection of about 120,000 black holes is generically described by solutions with $\phi_0 \in [4.25,5.5]$ and $\Phi_1/\Phi_{1,\rm max} \in [0.35,0.43]$. 

Using these near-critical black holes, we can systematically study the approach of various thermodynamic quantities to criticality. Since the behavior of these quantities is typically expected to be power law in the vicinity of the second order point, it is natural to study them on log-log plots on which critical exponents are trivially related to the slope of the best fit to the data. In practice, we extract this slope by performing a linear regression via a least squares fit. All reported critical exponents in this section have been obtained in this way.

\subsection{First-order line and critical density}

Despite knowing the location $(T_c, \mu_c)$ of the critical point, it remains a challenge to identify the critical density $\rho_c$.\footnote{Remarks about $\rho$ in this section apply equally well to $s$.}  Due to the infinite slope of the curve $\rho(\mu)$ on the critical isotherm, a large number of black holes with very different values of $\rho$ sit very close to the critical point (see the middle plot in figure~\ref{fig:rhoMu}).

We can  calculate the critical density in the process of determining the exponent $\beta$ which measures the rate that the discontinuity $\Delta \rho$ across the first-order line goes to zero as the critical point is approached:
\eqn{BetaAgain}{
\Delta \rho \sim (T_c - T)^\beta \,, \quad \quad \quad {\rm along \; first \; order \; line} \,.
}
In order to determine the discontinuity in $\rho$ for a given $T_f < T_c$, we must identify the $\mu_f$ at which the true ground state of the system jumps from the lower branch to the upper branch; this is the $\mu_f$ for which the free energy is the same for the two branches.  However we have not calculated the free energy, so we cannot identify $\mu_f$ in this way.  Equivalently, one can use Maxwell's equal-area construction, which states that $\mu_f$ should be placed such that the closed regions bounded by the isotherm on either side of the $\mu = \mu_f$ line are of equal area.

\begin{figure}
\begin{center}
\includegraphics[scale=0.35]{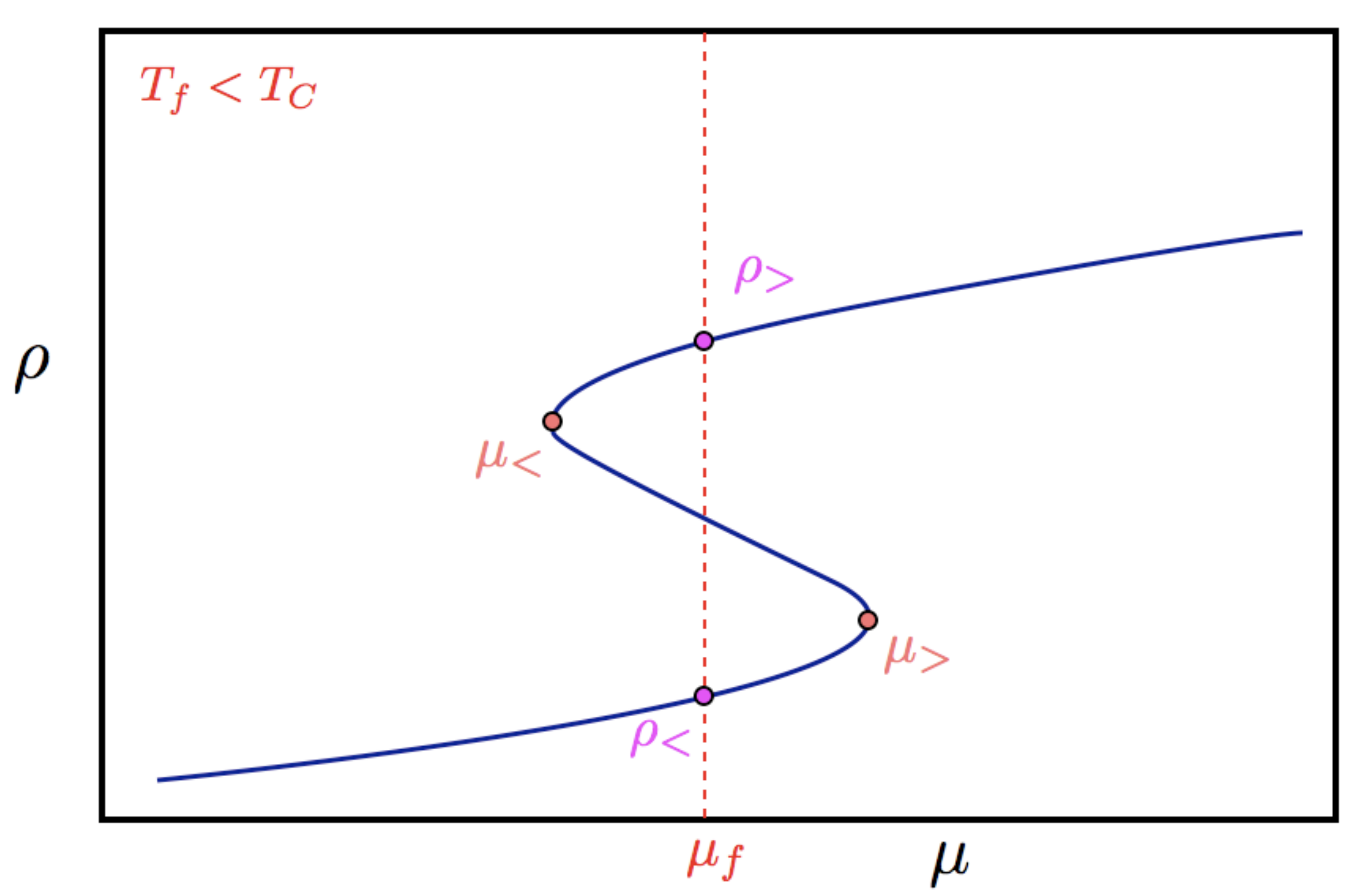}
\caption{Cartoon of $\rho(\mu)$ for  an isotherm with $T_f < T_c$, showing multivaluedness near the first-order line.  At the location of the line $\mu = \mu_f$ the true minimum of the free energy jumps from the lower to the upper branch and $\rho$ is discontinuous.
\label{fig:mixRho}}
\end{center}
\end{figure}

 We chose a computationally easier procedure which is asymptotically equivalent to the equal-area law as one approaches the critical point.  Namely, at a fixed temperature $T=T_f$, we define $\mu_<$ and $\mu_>$ to be the locations of the local minimum and maximum of the isotherm $\rho(\mu)$, and we define $\mu_f$ 
to be the midpoint between them.   This in turn determines the mixed-phase densities $\rho_<$ and $\rho_>$ for the point $(T_f, \mu_f)$ along the first-order line.  This procedure is illustrated in figure \ref{fig:mixRho}.
For points we checked near the critical point, this procedure agrees with the equal-area rule to within a fraction of a percent.

\begin{figure}
\begin{center}
\includegraphics[scale=0.35]{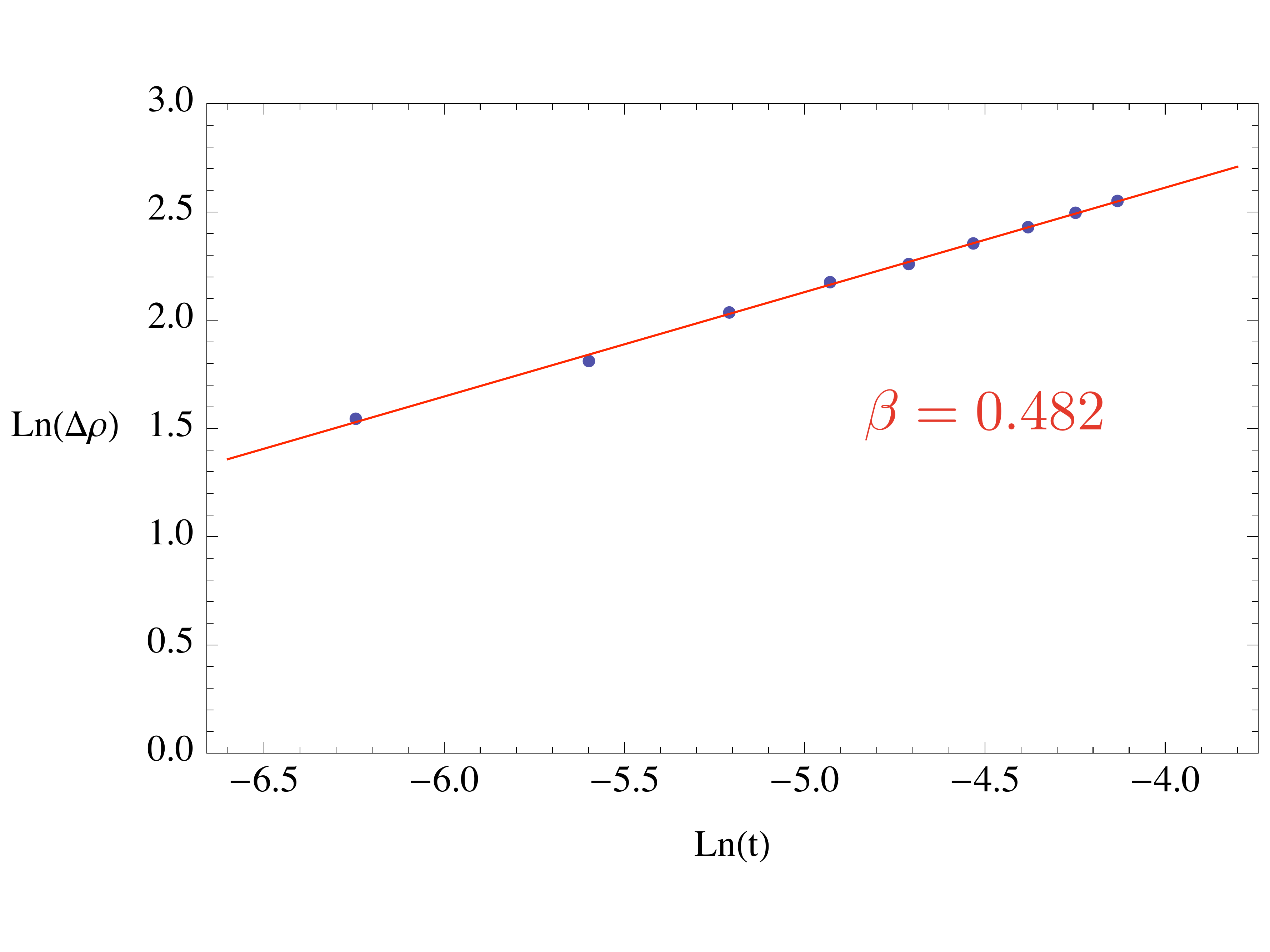}
\caption{The discontinuity in the baryon density as the critical point is approached on a log-log plot.  The slope of a best fit line through the data gives us a value $\beta = 0.482$.
\label{fig:beta}}
\end{center}
\end{figure}

The critical density $\rho_c$ is then most easily obtained as the limit that both $\rho_<$ and $\rho_>$ approach as we near the critical point.  The result is
\eqn{RhoC}{
\rho_c = 9.9022 \,.
}
Plotting $\Delta \rho \equiv \rho_> - \rho_<$ in a log-log plot with $t \equiv (T - T_c)/T_c$, we obtain
\eqn{BetaResult}{
\beta \approx 0.482 \,.
}
In comparison, the  exponent in the mean-field case is $\beta_{MF} = 1/2$, and so we have found a result very close to the mean-field value.

\subsection{Critical isotherm}

\begin{figure}
\begin{center}
\includegraphics[scale=1.2]{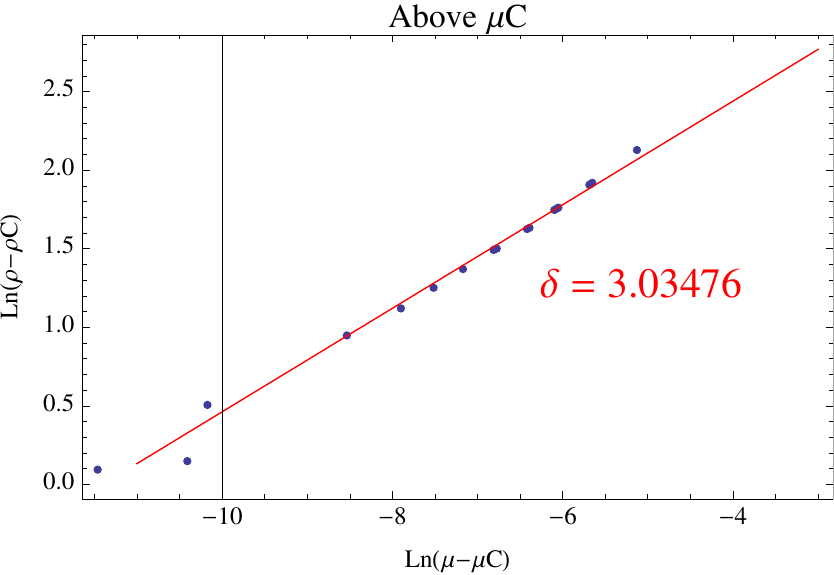}
\caption{The rate at which $\rho$ approaches $\rho_c$ as $\mu$ approaches $\mu_c$ on the critical isotherm on a log-log plot.  The slope gives us a value $\delta = 3.03476$.
\label{fig:delta}}
\end{center}
\end{figure}

As discussed in the previous section, the critical isotherm at $T = T_c$ is the curve marking the boundary between single-valued and multi-valued behavior, and correspondingly it has a diverging slope for $\rho$ and $s$ precisely at $\mu = \mu_c$. Using the behavior of $\rho$ on the critical isotherm, we can determine the critical exponent $\delta$, defined as
\eqn{DeltaAgain}{
\rho - \rho_c \sim |\mu - \mu_c|^{1/\delta} \,, \quad \quad \quad {\rm for}\ T = T_c \,.
}
Now that we have $\rho_c$ in hand, we plot a number of black holes on a log-log plot near the critical point with $\mu > \mu_c$ in figure~\ref{fig:delta}.  The  points are fit well by a straight line with slope giving $\delta = 3.03476$.  We note that the mean field value is $\delta =3$.

\subsection{First order axis and susceptibilities}\label{sec:alpha}

\begin{figure}
\begin{center}
\includegraphics[scale=0.3]{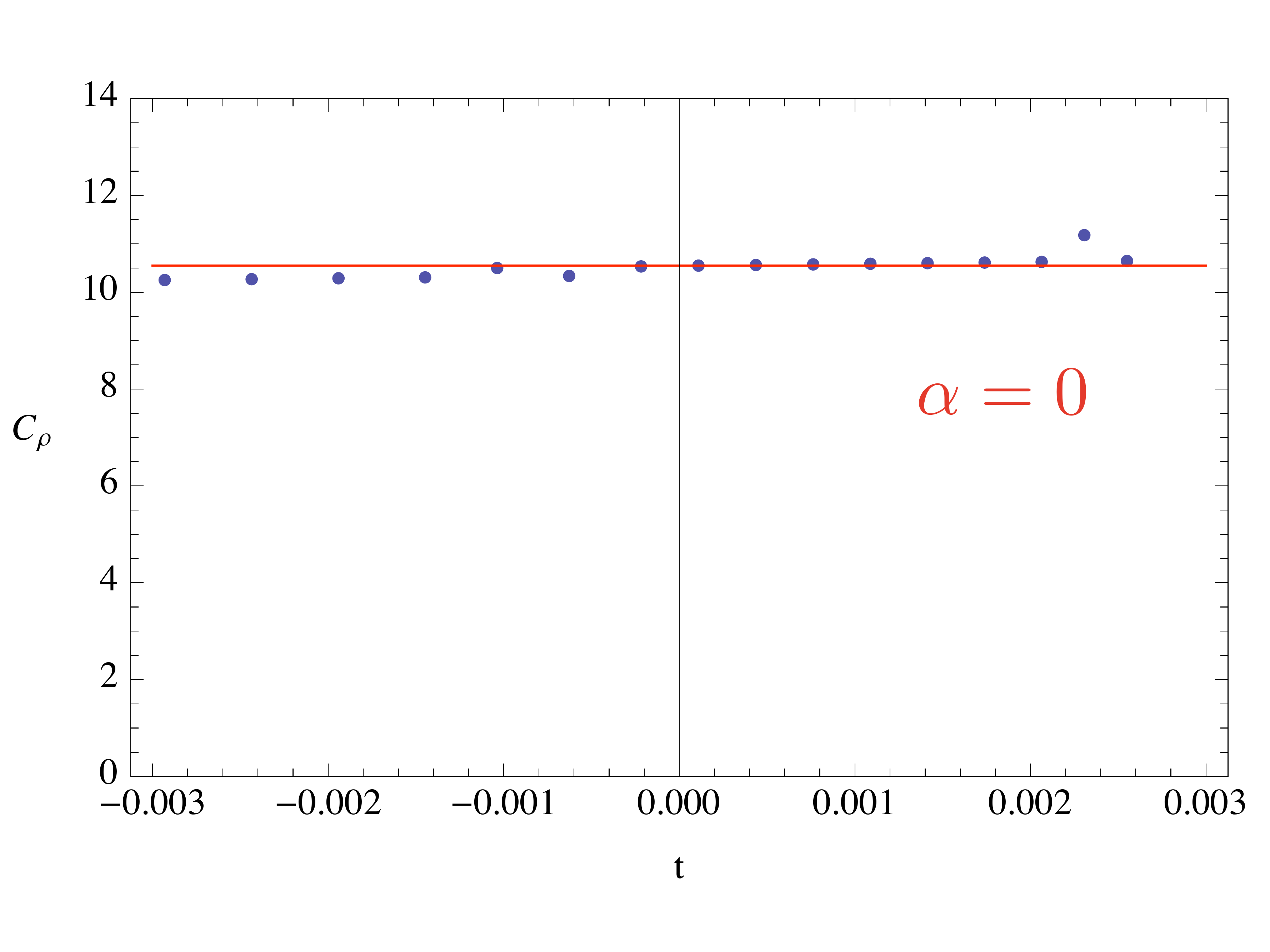}
\caption{The specific heat $C_\rho$ near the critical point along a line of constant $\rho$, along the first-order axis.  There is no divergence, giving $\alpha = 0$.
\label{fig:alpha}}
\end{center}
\end{figure}

Susceptibilities in general diverge near a critical point.  However, it is not required that all possible susceptibilities are divergent.  Indeed we find this is the case for $\alpha$, which is the power law exponent for the specific heat $C_\rho$ along the axis defined by the first order line:
\eqn{AlphaAgain}{
C_\rho \sim |T - T_c|^{-\alpha} \,, \quad \quad \quad {\rm along \; first \; order \; axis} \,.
}
To avoid the complications of the first-order line itself, we perform this approach from the other side, with $\mu < \mu_c$.  A calculational advantage is that the line of constant $\rho$ comes very close to the first order axis \cite{Griffiths:1970zz}, so we can simply generate a set of black holes filling out that line, and define $C_\rho$ in terms of finite differences of nearest neighbors.

The result is that $C_\rho$ does not diverge at all along this line, but instead is smooth at values near $C_\rho \approx 10.5$.  This corresponds to a vanishing $\alpha$:
\eqn{AlphaResult}{
\alpha =0 \,.
}
In a sense this is the most robust of all our results, since even a weak divergence looks completely different from a lack of divergence; it suggests that the result \eno{AlphaResult} is exact.  Moreover, this is again the mean field result; for example in the van der Waals theory of a fluid one has $C_\rho = 3n/2$ with $n$ the number density.\footnote{Certain systems contain a discontinuity or logarithmic divergence for $C_\rho$, and are also grouped as $\alpha = 0$; ours is truly smooth, as with the van der Waals field.}
\begin{figure}
\begin{center}
\includegraphics[scale=1.1]{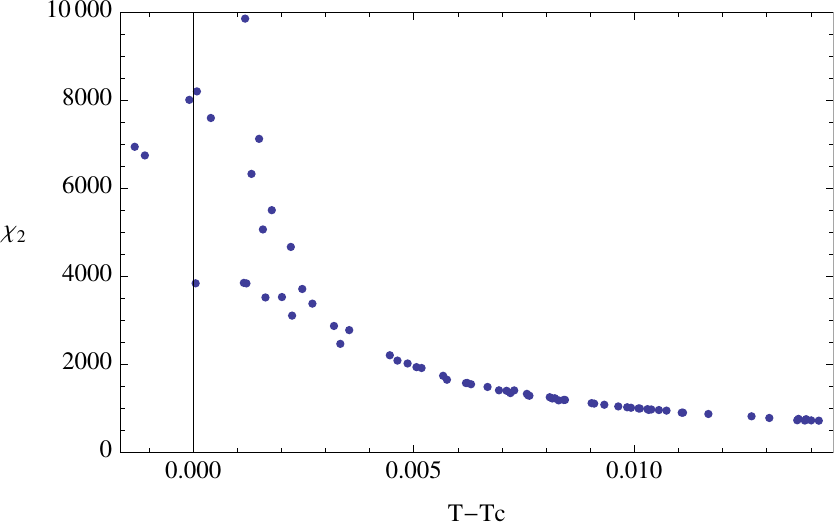}
\caption{The susceptibility $\chi_2$ near the critical point along a line of constant $\rho$.
\label{fig:gammanolog}}
\end{center}
\end{figure}

\begin{figure}
\begin{center}
\includegraphics[scale=0.3]{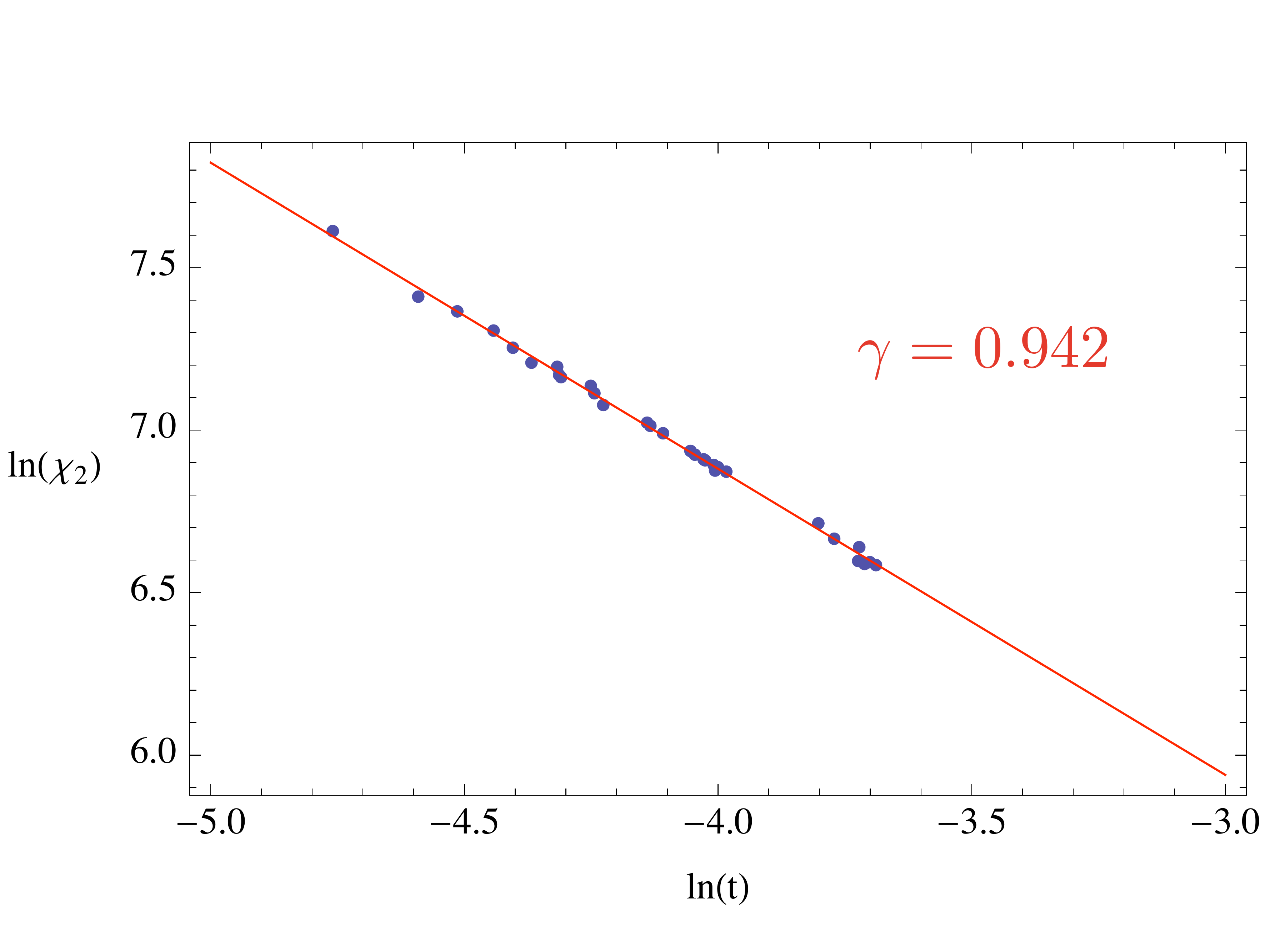}
\caption{The baryon susceptibility $\chi_2$ compared to $t \equiv (T - T_c)/T_c$ as the critical point is approached on a log-log plot.  The slope gives us a value $\gamma = 0.942$.
\label{fig:gamma}}
\end{center}
\end{figure}

Although we find $C_\rho$ to have no divergence, other quantities do show the expected divergences.  The final thermodynamic exponent, $\gamma$, is defined by the approach of $\chi_2$ along the same axis,
\eqn{GammaAgain}{
\chi_2 \sim |T - T_c|^{-\gamma} \,, \quad \quad \quad {\rm along \; first \; order \; axis} \,.
}
This exponent is the most difficult to calculate.  The quantity involves a derivative in the $\mu$-direction, so to calculate the finite difference we must obtain pairs of points with the same $T$ to within a very small tolerance $\Delta T$, which are separated by a larger but still small amount $\Delta \mu$; we then need a sequence of pairs of such points moving along the first-order axis, a direction unrelated to the derivative.  To find black holes near the axis, we again imposed constant $\rho$.
We then looked for pairs of points  with $\Delta \mu < 0.001$ and 
kept those with $\Delta T / \Delta \mu < 0.002$.

The result is shown in figure~\ref{fig:gammanolog}.  The black holes for $T - T_c > 0.004$ form a smooth, single-valued curve, but near the critical point numerical errors grow larger and the curve is no longer single-valued.  Fitting just the single-valued region, we produce the log-log plot in figure~\ref{fig:gamma}.  The resulting exponent is $\gamma = 0.942$, while
the mean field value of $\gamma$ is $\gamma_{MF} = 1$.  Once again our result is consistent with mean field.

\subsection{Summary and Scaling}\label{sec:critExpConcs}

In conclusion, we have measured the critical exponents $\alpha$, $\beta$, $\gamma$ and $\delta$ and found them to be consistent with the mean field values.  Since the mean field values are themselves consistent with scaling \eno{Scaling}, it is clear that our results pass this self-consistency check as well.

As an idea of the size of the errors in our measurements, we can choose two exponents, use the scaling laws \eno{Scaling} to calculate predictions for the other two, and compare these predictions to our actual results.  It is natural to use $\alpha = 0$ as an input, since we obtain it as an apparently exact result.  Also inputting $\delta = 3.035$  gives us the results:
\begin{center}
\begin{tabular}{|c|c|c|c|c|}
\hline
Exponent &$\alpha$ & $\beta$ & $\gamma$ & $\delta$ \\
\hline
Calculated value & 0 & 0.482 & 0.942 & 3.035 \\
\hline
Scaling prediction from $\alpha$ \& $\delta$ & 0 & 0.496 & 1.009 & 3.035 \\
\hline
\% Diff. &--- &3\% &7\%  & --- \\
\hline
\end{tabular}
\end{center}
Thus our deviations from scaling are in the $3\%-7\%$ range, giving an idea of the size of the errors in our method.

\subsection{Other models}

We also carried through an analysis of the model with the same potential \eno{VChoice} and the gauge kinetic function
\eqn{fExp}{
f(\phi) = e^{-\phi} \,.
}
The phase diagram obtained from this model shares all relevant properties to the one presented here: a first-order line ending in a critical point, and critical exponents consistent with mean field.  We omit the details since they are virtually identical to those just discussed. This exponential model has substantially poorer fit to lattice data at $\mu=0$.

\section{Conclusions}
\label{CONCLUSIONS}

A number of techniques have been employed to predict the location of the QCD critical point.  These include lattice calculations that attempt to circumvent the problems of finite $\mu$ by a number of different means, including taking a Taylor series expansion around $\mu =0$, reweighting the contributions to the path integral, or analytically continuing from imaginary chemical potential.  There are also calculations in a variety of Nambu--Jona-Lasinio models, along with a number of other methods.

\begin{figure}
\begin{center}
\includegraphics[scale=0.3]{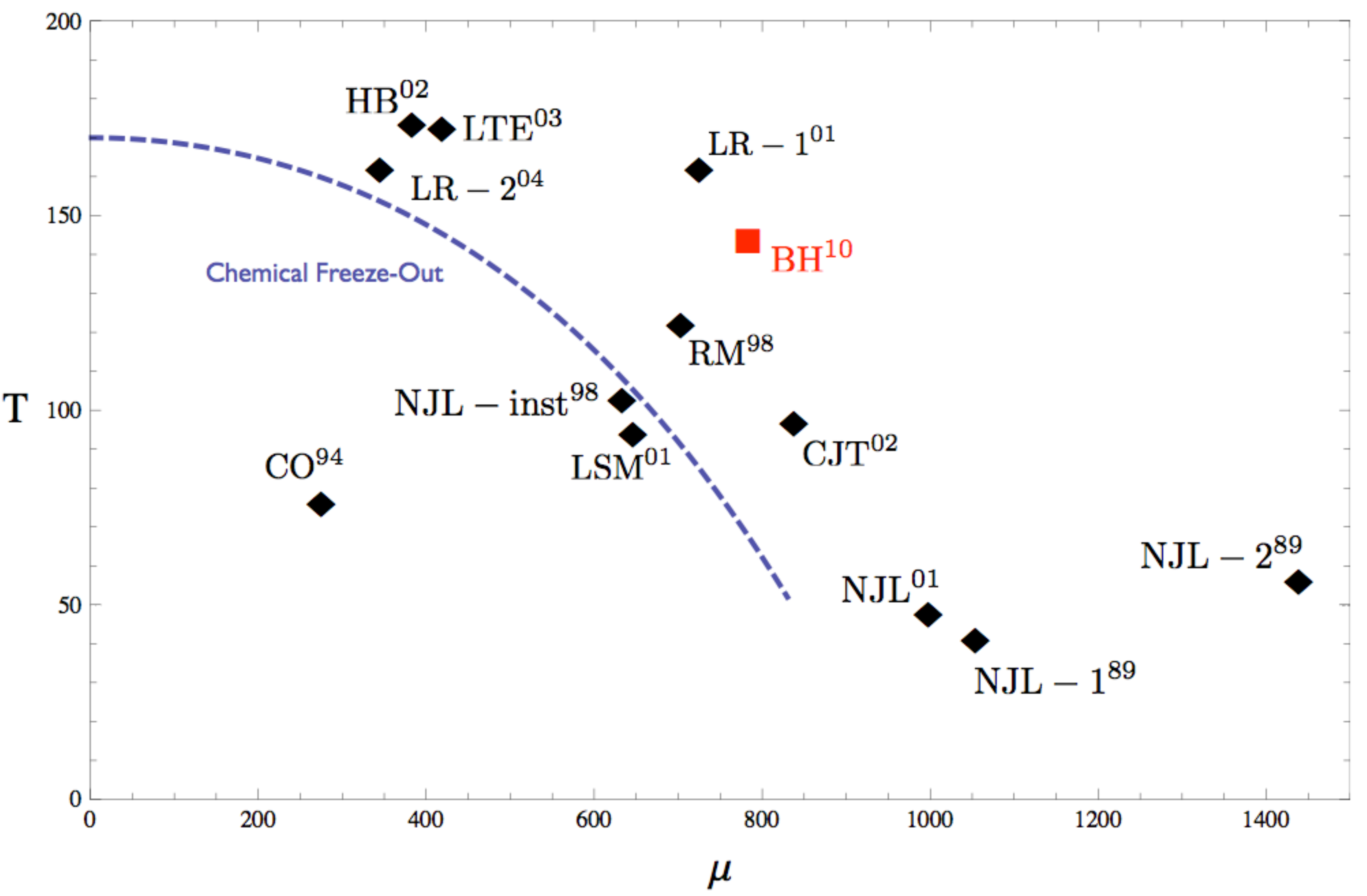}
\caption{Our result for the location of the critical point (BH$^{10}$) compared to other calculations in the literature, along with the chemical freeze-out curve. 
\label{cpPredsPlot}}
\end{center}
\end{figure}

In figure~\ref{cpPredsPlot}, we show the location of a number of different calculations of the location of the critical point, along with our result, presented as BH$^{10}$.  A key to the various abbreviations and references is given in the table; for more information see \cite{Stephanov:2004wx,Stephanov:2005iu}.   The variation in prior results is considerable, and our result lies within the parameter space defined by the others.
Also included in this plot is an estimate for the chemical freeze-out line \cite{Aggarwal:2010cw}.   

\begin{center}\begin{tabular}{|c|c|c|}
\hline
Label & Method & Reference \\
\hline
\hline
HB & Hadronic Bootstrap & \cite{Antoniou:2002xq} \\ 
\hline
LTE & Lattice Taylor Expansion & \cite{Ejiri:2003dc}\\
\hline
LR & Lattice Reweighting & \cite{Fodor:2001pe}, \cite{Fodor:2004nz} \\
 \hline
RM & Random Matrix & \cite{Halasz:1998qr}\\
\hline
NJL & Nambu--Jona-Lasinio & \cite{Asakawa:1989bq},\cite{Scavenius:2000qd},\cite{Berges:1998rc}\\ 
\hline
CJT & Effective Potential & \cite{Hatta:2002sj} \\
\hline
LSM & Linear Sigma Model & \cite{Scavenius:2000qd} \\
\hline
CO & Composite Operator &\cite{Barducci:1989wi}, \cite{Barducci:1993bh} \\
\hline
\end{tabular}\end{center}

\medskip
\noindent
In general, as heavy ion collisions gain center-of-mass energy, the produced medium is characterized by higher $T$ and lower $\mu$.
This leads to several issues with the possibility of exploring the region near our prediction, where $\mu$ is relatively large.  First, assuming the heavy ion collisions attain thermodynamic equilibrium, the value of $T$ that can be reached may be too small by the time one has reached sufficiently large $\mu$.  Moreover, at some large $\mu$ the center-of-mass energy will become too low to actually thermalize the colliding ions, making a thermodynamic interpretation no longer appropriate.  Collisions at RHIC have a minimum energy of 5-7 GeV \cite{Aggarwal:2010cw}, too high to reach our value of $\mu_c$; LHC is even worse.  A more promising possibility is the CBM experiment at the future accelerator center FAIR at GSI, a fixed-target experiment whose intended region of exploration includes the location of our critical point \cite{Staszel:2010zz}.

That said, as we have emphasized our result should be regarded primarily as a proof of principle: it is possible to extract QCD-like phase diagrams from relatively simple holographic duals.  This result has substantial promise, precisely because finite chemical potential calculations are so difficult on the lattice; in gravity duals, finite chemical potential simply involves introducing a new field, and possesses no additional qualitative complexity.  Of course holographic duals introduce other complications---large $N$ and the fact that they are not precisely QCD, only QCD-like---that must in turn be dealt with.

We have discovered a first-order line and a critical endpoint with mean-field critical exponents.  A number of ways to generalize these results are evident.  Most obviously, one can study the fluctuations around our classical backgrounds, thereby learning about spectra, transport and the true free energy function.  An equally obvious, though potentially difficult, further step is to add $1/N$ corrections to the geometries, hopefully moving the critical point away from mean field.  One can also consider studying a larger theory.  In particular, the introduction of chiral symmetry in addition to baryon number is straightforward in principle, if more intricate in practice.  By enlarging the theory one could also hope to study the color superconducting phases at large $\mu$. We hope to examine these issues in the future.


\section*{Acknowledgments}

We are grateful to Tom DeGrand, Victor Gurarie, Anna Hasenfratz, Mike Hermele, Jamie Nagle, Leo Radzihovsky, Paul Romatschke and Dam Son for helpful discussions.
The work of O.D.\ and C.R.\ was supported by the Department of Energy under Grant
No.~DE-FG02-91-ER-40672.  The work of S.S.G.\ was supported by the Department of Energy under Grant No.~DE-FG02-91ER40671.

\bibliographystyle{JHEP}
\bibliography{phase}

\end{document}